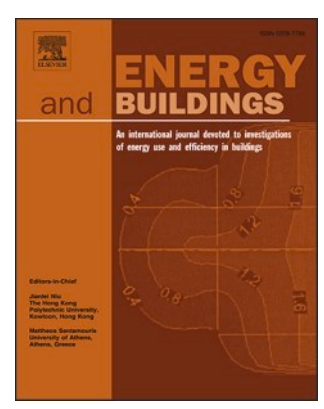

# Carbon reductions through optimized solar heat gain glass properties considering future climate and grid emissions: case study of Chicago's residential buildings

Yiwei Lyu [a,b], Jialiang Xiang [a], Holly Samuelson [a,b,*]

[a] Graduate School of Design, Harvard University, Cambridge, MA 02138, USA
[b] Harvard Center for Green Buildings and Cities, Harvard University, Cambridge, MA 02138, USA

ABSTRACT

Existing resources leave confusion over the benefits of high versus low Solar Heat Gain Coefficient (SHGC) windows for energy performance in residential buildings retrofits in cold climates. Additionally, few studies have considered the impact of expected future climate conditions and time-variable grid emission rates on energy-related metrics. Utilizing the ResStock, residential building stock models from the National Renewable Energy Laboratory (NREL), this study investigates retrofits increasing the SHGC of windows in Chicago, a cold US city. The results indicate that increasing window SHGC increases summer cooling needs; however, in most cases, this effect is more than offset by reduced winter heating needs. This balance is particularly beneficial considering the state's expected long-run marginal carbon emission rates. The study also examines the combined effects of high SHGC with improved window insulation values, demonstrating that such strategic window retrofits not only enhance overall building energy performance but also contribute to greater emission reductions. On average, the current Chicago residences (n = 4,826) save 4.6 % on heating and cooling carbon emissions by increasing the SHGC of the windows. If we assume that those homes are upgraded with heat pumps (electrification), a popular retrofit that reduces heating-related carbon emissions in particular, the increased window SHGC saves 2.5 % of long-run marginal carbon emissions. These results provide new insight into the carbon benefits of higher SHGC replacement windows in a cold climate. The benefits are significant, even considering future trends of a warming climate, higher demand grid emissions, and building electrification.

## 1. Introduction

As concerns about climate change escalate, enhancing the energy efficiency of buildings has become critical in reducing greenhouse gas emissions [1]. This paper explores an under-researched approach to boost building energy efficiency and mitigate emissions by focusing on the Solar Heat Gain Coefficient[1] (SHGC) of windows in existing residential buildings, particularly in a cold-winter, hot-summer climate where the net-benefit versus net-detriment of passive solar heat gain is not universally agreed upon. The relevance of this study is amplified by the ongoing transition towards the electrification of buildings [3], a shift anticipated to substantially increase heating demands on the grid [4] particularly in colder climates like Chicago [5]. This transition will not only challenge the electricity infrastructure in cold weather [4] but also presents a unique opportunity to redefine and improve the energy consumption patterns in buildings for enhanced sustainability and resilience.

In the evolving landscape of building energy efficiency, the optimal SHGC for windows remains a topic of debate. For new buildings and major retrofits, the prescriptive path of U.S. building standards and codes, including ASHRAE 90.1/90.2 [6,7] and the International Energy Conservation Code (IECC) [8], establish a specific *maximum* on window SHGC, i.e. promoting lower SHGC values even in residential buildings in cool climates, like Chicago.[2] These building standards favor lower SHGC presumably because they are designed for new constructions and major retrofits with more insulative envelopes and potentially separate heating and cooling systems, possibly aiming to reduce cooling loads and allow for smaller cooling systems. In contrast, EnergyStar, which certifies high-performing residential windows,[3] sets a *minimum* on SHGC in colder climates, i.e. promoting higher SHGC values to enhance energy conservation [10]. Online resources such as the Efficient Windows Collaborative, supported by the National Fenestration Rating Council (NFRC), provide architects and builders with tools and guidance to select windows that optimally balance SHGC with other critical factors like U-value and visible light transmission [11]. When tested with a Chicago zip code, the tool recommended windows with a high SHGC of 0.5 [11].

* Corresponding author.
*E-mail address:* hsamuelson@gsd.harvard.edu (H. Samuelson).

[1] The Solar Heat Gain Coefficient (SHGC) measures the unitless fraction of solar radiation admitted through a window, both directly transmitted and absorbed, then released as heat inside a building [2].

Outside the US, building codes in countries like India and China mirror the US in requiring low SHGC values to mitigate excessive solar heat gain in warm and hot climate regions. However, they differ from the US for colder regions like ASHRAE Climate Zone 5. In similar cold climates, these international standards allow more solar heat gain. For example, while ASHRAE 90.1,90.2 and IECC [6,7,8] in the US set a maximum SHGC of 0.33 for multi-family residential buildings over three stories (and 0.4 for other residential buildings), in cold B climate in China—Similar to ASHRAE Zone 5—the maximum allowable SHGC is 0.55 [12]. Moreover, the Chinese SHGC requirement is orientation-specific, only affecting windows facing west or east [12]. This contrasts with the uniform SHGC value applied across all orientations in the U.S. codes. Similarly, in India, the GRIHA 2019 code mandates a maximum allowable SHGC of 0.62 for colder climates [13]. These Chinese and Indian maximum SHGC limits are significantly higher than the prescriptive U.S. maximum allowable SHGC similar cold climate zones.

Scholarly studies, such as those conducted by the National Renewable Energy Laboratory (NREL), provide practical insights by simulating the energy-related performance of different residential window properties. However, readers looking for “good, better, and best” SHGC values may be confused, because in such studies’ window selections [14], the SHGC decreases as the U-values improve. This trend may result from strategies like adding an extra windowpane or applying a low-e coating to reduce the U-value, which often also lowers the SHGC [15,16]. It is well-established that enhancements in window U-value decrease heat transfer through windows, thereby reducing the energy needed for both heating and cooling [14]. Similarly, the analysis of the 2019 update of ASHRAE Standard 90.1 highlights energy reductions from the older to the newer version of the standard [17]. Notably, this update to the standard lowered cold climate SHGC requirements from 0.38 to 0.33. Therefore, this study demonstrated that the “better” performing cases had lower SHGC windows, but, again, the U-values of the windows were also improved. Therefore, although in such studies, the “lower” and “lowest” SHGC windows tested exhibit the “better” and “best” energy performance, one cannot infer that the change in SHGC was the cause of such improvements.

Meanwhile, our review of product catalogs from various glass manufacturers reveals that this relationship (lower U-value always coinciding with lower SHGC) is not consistent across all window products [18,19]. We observed numerous instances where windows with low U-values also exhibited high SHGC, indicating that these properties can vary somewhat independently depending on the glass selection and coatings or infill gases used. For specific window examples, please refer to Appendix A.

This body of literature and resources described above underscores the ambiguity surrounding the optimal SHGC values for window retrofit projects in cold climates. It highlights the need for ongoing research and adaptable, context-specific standards in window design, ensuring that retrofit teams can effectively target and test SHGC values that best meet the diverse energy performance goals of different buildings and locations.

Moreover, to our knowledge, these resources do not yet account for estimations of future climate conditions or the carbon emission profiles of different electricity grid locations. Such an analysis could provide insight to help achieve long-term sustainability goals. Long-run marginal emission rates (LRMER) measure the future change in emissions from each unit of electricity used or saved [20]. Because LRMER considers how the change could influence both the operation and structure of the grid, it estimates more accurately for long-term interventions [20]. Understanding these emissions is helpful for assessing the environmental impact of energy-saving strategies over time [21], and enables users to consider building design decisions in the context of future states of the energy grid and climate conditions [22].

To address these mentioned gaps, our research utilizes NREL ResStock models [23,24], which we have adapted to simulate the impact of increased SHGC and incorporate expected future climate conditions. ResStock is a suite of energy models designed to accurately represent the existing residential building stock and have been thoroughly validated by NREL [25,26]. To account for future climate shifts, the use of morphed weather files is a widely recognized and accepted practice in the field of building science [27,28]. Morphed weather files are modified versions of existing meteorological data, adjusted to reflect anticipated changes in climate conditions such as temperature, humidity, and precipitation patterns [29]. These files are helpful for simulating how buildings will perform under future climatic scenarios [30]. The use of these future climate files is helpful, since windows selected today are likely to remain in place as climates shift, impacting building performance and energy requirements.

These modeling techniques allow us to quantify the potential energy savings and emission reductions achievable through strategic building design modifications. Additionally, the study involves a detailed data analysis, using these results to identify key architectural features within the building stock that significantly impact potential energy savings associated with variations in SHGC.

Central to our analysis is the hypothesis that increasing the SHGC can crucially balance building energy demands throughout the year. While higher SHGC values are likely to increase cooling requirements in the summer, they can substantially decrease heating needs during the winter [14]. This dynamic is particularly significant in climates like Chicago’s, where winter heating significantly contributes to annual energy usage [31]. Furthermore, as the electrical grid evolves towards cleaner energy sources [3] —especially during summer when solar power production is at its peak—the proposed window modifications could better align building energy demands with periods of lower-emission electricity generation, thereby reducing long-run marginal emissions.

Our methodology also examines the effect of enhancing thermal insulation through improvements in the U-value of windows, alongside adjustments to SHGC. By evaluating scenarios that combine high SHGC with improved U-values, our study extends the NREL ResStock models to provide a comprehensive analysis of how these dual modifications can synergistically boost building energy and carbon emission performance.

[2] Residential buildings under three stories are governed by residential requirements, while buildings taller than three stories are subject to the commercial provisions of these codes. Building codes apply primarily to new constructions and major renovations. They offer both prescriptive and performance-based compliance paths, allowing project teams to opt for non-standard SHGC values by demonstrating, through energy modeling, provided that their building’s overall performance meets or exceeds that of a code-compliant building.

[3] This requirement is relevant to building owners as they can benefit financially from installing EnergyStar windows, with tax credits of up to 30% of the product cost, up to $600 [9].

This combined approach is crucial, as it addresses the complex interplay between heat gain and insulation, optimizing the overall energy efficiency of buildings throughout the year.

Through this comprehensive methodology, our study aims to not only demonstrate the individual and combined impacts of SHGC and U-value modifications but also to illuminate the broader benefits these changes could offer to urban buildings in terms of energy efficiency and carbon emission reductions. These findings are intended to inform and assist stakeholders in retrofitting residential buildings and the policies that influence them, fostering a move towards more sustainable and resilient urban landscapes.

## 2. Methods

### 2.1. Hypothesis testing

Central to our investigation are two hypotheses that explore the role of window properties in optimizing the energy performance of buildings in cities with cold climates, using models representing Chicago's residential housing stock as a case study. First, we hypothesize that using south-facing windows with a higher SHGC than typically recommended by best practices and building codes can reduce overall emissions associated with heating and cooling. This hypothesis was informed by the potential for increased solar gain through south-facing windows, which might be more beneficial than gains through windows of other orientations [32]. Second, we posit that windows that combine improved insulation values with high SHGC will perform better in terms of long-run marginal emissions than those with improved insulation alone.

These hypotheses are tested by analyzing how varying levels of insulation and SHGC impact annual energy consumption and emissions in various building models, focusing particularly on the combined effects of these modifications in cold weather conditions. This approach will allow us to determine if the strategic selection of window properties can indeed optimize building performance.

### 2.2. Research Framework

The choice of Chicago as a focal point for this study is driven by its climatic characteristics, where winter heating demands are substantial. Chicago is in ASHRAE climate Zone 5A [5,33] characterized by long, cold winters, and hot, humid summers [34]. The study aims to demonstrate how increased SHGC can reduce heating demands in colder months, potentially offsetting increased cooling needs in warmer months, in net annual energy performance. Additionally, we consider the evolving nature of the energy grid, particularly the increasing incorporation of solar energy which peaks during summer months. The modifications suggested in this research could help align building energy demands with periods of lower-emission electricity generation, thereby optimizing energy use and reducing overall emissions.

In our research methods, we modified versions of the NREL 2022 ResStock models. Specifically, we adapted the models to simulate the effects of varying SHGC. As shown in Fig. 1, the emission rates we used for the simulation are the long-run marginal emission rates of Illinois from the 2022 NREL Cambium data [20]. As illustrated in Fig. 1, Cambium emissions are not constant[4]; they tend to be lower during the mild spring and fall months and increase during the more extreme hot and cold months. MRI-ESM2.0 2020–2039 weather file for Chicago [36] was selected for its moderate prediction among the six climate models presented by Bass et al. [28]. This weather file has a higher monthly average temperature than the typical TMY file with more severe weather days. The primary aim of this research is to evaluate the potential energy savings and corresponding reduction in emissions that can be achieved through strategic enhancements in residential window replacement selection.

### 2.3. Model modification and simulation

To achieve our objectives, we modified the ResStock models to incorporate different SHGC values in the simulation environment. ResStock [37] utilizes OpenStudio® [38] and EnergyPlus™ [39] within the open-source building energy modeling ecosystem of Department of Energy (DOE). Combining housing stock characteristics database, physical-based computer modeling, and high-performance computing [25], ResStock models have been extensively validated by NREL to ensure accuracy and reliability in predicting energy usage and potential savings in residential buildings [26].

ResStock includes representative housing types. For example, in Chicago this includes single family homes, apartment units in multifamily housing, etc. Fig. 2 visualizes the different residential building typologies included in our simulations. To understand the existing condition as well as future electrified scenarios, we designed two studies with four sets of results for both ResStock baseline cases[5] (labeled as 1A and 2A) and ResStock heat pump upgrade cases[6] with the existing heating system used as backup (labeled as 1B and 2B). These two sets of ResStock models were chosen for specific reasons: the 2018 building stock models accurately represent the current state of buildings [26], while the heat pump upgrade cases reflect anticipated near-future trends [41]. Heat pumps are increasingly recognized as a cost-effective method for saving energy and are being adopted widely [41], making them a relevant focus for future-oriented energy modeling.

We ran two sets of 2,176 cases for the first study and two sets of 4,826 instances for the second study (detailed description below). The difference in the number of cases is due to the specific focus of Study 1, which exclusively examined buildings with south-facing windows.[7] Consequently, we applied an additional filter to only include buildings oriented in a north–south direction, resulting in a smaller dataset compared to Study 2, which did not have this orientation constraint.

#### 2.3.1. Study 1: Evaluating the impact of high SHGC for South-Facing windows

In Study 1, we focused on evaluating the effects of significantly increasing the SHGC of south-facing windows. The average SHGC of the 2,176 cases studied is 0.57. For this study, we tested the impact of increasing the south-facing window glass to SHGC to 0.7, in ResStock cases where it was lower to start. This 0.7 SHGC value is considerably higher than the maximum values allowed by ASHRAE 90.1 (0.33) and IECC standards (0.4) [6,8]. However, we hypothesize that the decreased heating load, due to passive solar heating, will offset any increase in cooling load. This 0.7 SHGC selection was based on a review of existing products in manufacturers' catalogs [18,19], ensuring that there are commercially available window options that not only feature this higher SHGC but also maintain an adequate U-value, compliant with building codes for Chicago's climate zone [6].

To simulate and analyze the energy dynamics and emission impacts under enhanced SHGC, we conducted two distinct sets of simulations for Study 1A and Study 1B using the modified NREL ResStock models as follows.

[4] All Cambium data is created based on 2012 weather patterns. This study utilized the month-hour temporal aggregations of Illinois to capture the season variation while removing the weather-driven patterns, as recommended by the Cambium data documentations for weather alignment [35].

[5] Representing the 2018 residential building stock in Chicago.

[6] Centrally ducted single-speed heat pump, SEER 15, 9 HSPF. For more details, readers can refer to the ResStock documentation [40].

[7] We defined south-facing windows as windows with azimuth (orientation) angles between 165.0 and 195.0 degrees. Since all ResStock model windows are in one of eight directions (45 degrees increment), we only modified windows facing south for Study 1.

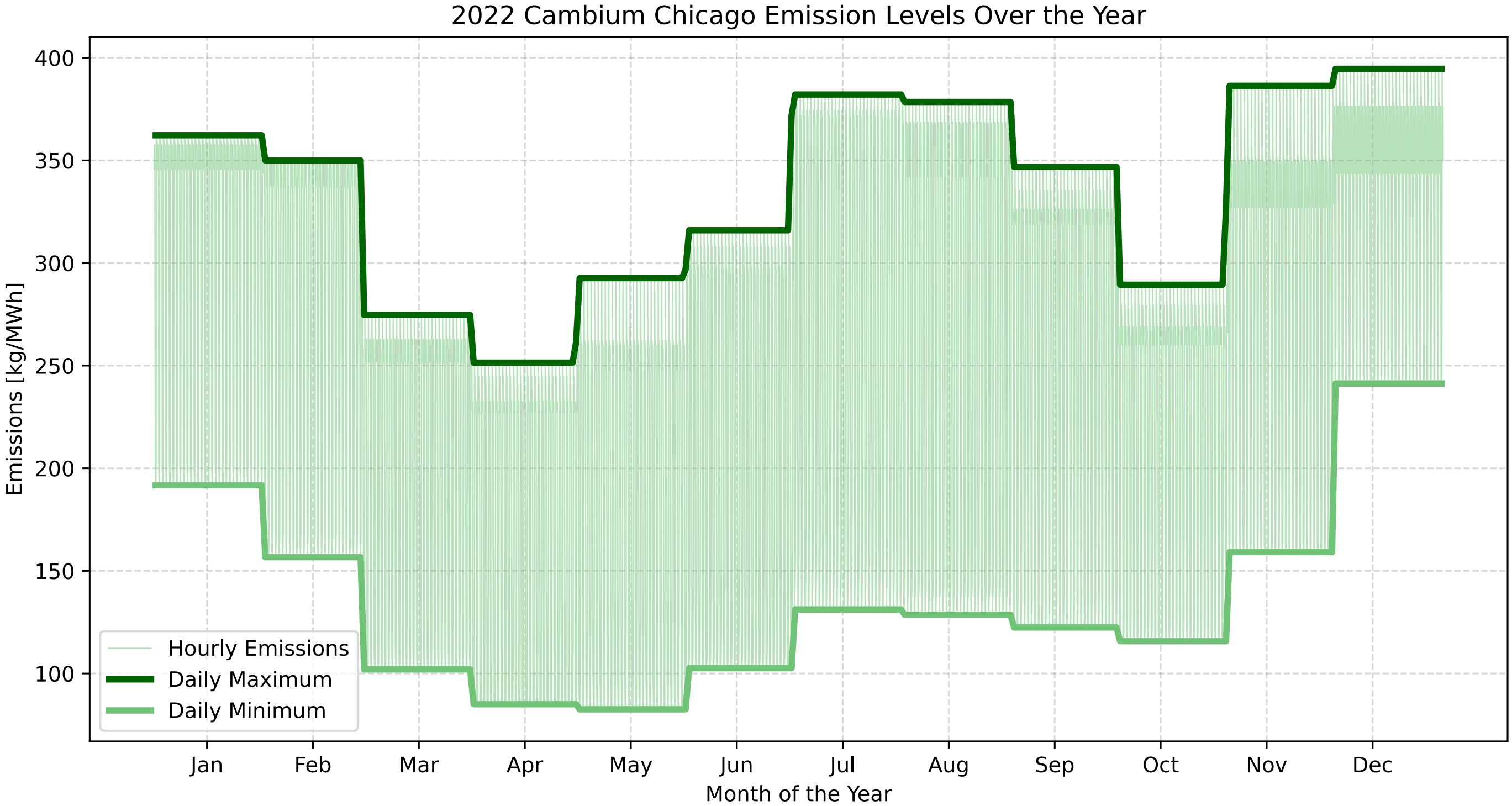

**Fig. 1.** Cambium long-run emission 2022, published in 2023 [20].

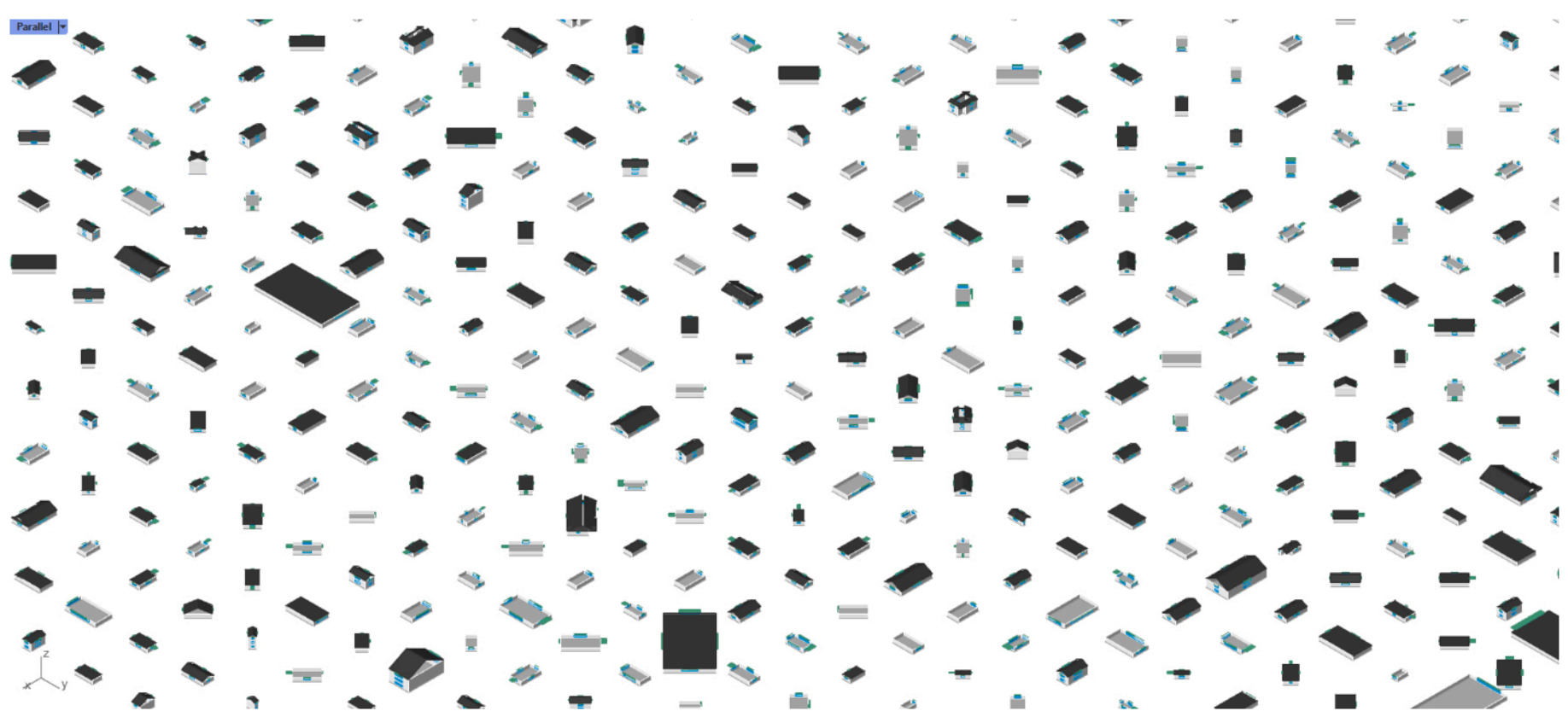

**Fig. 2.** Grasshopper visualization of Chicago ResStock models simulated.

1. Baseline Simulation: The first set of simulations was run using the updated climate and emission files as described in Section 2.1, without any alterations to the SHGC values. This simulation served as a control to gauge the baseline energy performance and emissions output of residential buildings in Chicago.
2. Modified SHGC Simulation: The second set involved identical climate and emission files but included the increase of SHGC for south-facing windows to 0.7. This approach allowed us to isolate the effects of increased solar heat gain from south-facing windows on the overall energy consumption and emissions of buildings.

#### 2.3.2. Study 2: Evaluating the combined effects of U-value and SHGC upgrades

Because it may be unrealistic to replace windows without taking the opportunity to improve their insulation value, Study 2 extends our examination of window properties by exploring the relationship between U-value upgrades and SHGC adjustments. This study tests various combinations of these factors to determine their collective impact on building energy efficiency and emissions. The average U-value of the 4,826 cases is 3.92 W/m$^2$·K (0.69 Btu/h·ft$^2$·F), and the average SHGC is 0.60. The specific upgrades, as outlined in Table 1, are designed to systematically evaluate how different levels of insulation and solar gain influence building performance in Chicago's climate. We intend to explore the potential benefits of retrofitting windows with enhanced properties. Therefore, we are selectively upgrading only those windows that do not meet existing performance requirements, while windows that already fulfill these standards remain unchanged. This targeted approach allows us to focus on the improvements directly attributable to the retrofit interventions.

**Table 1**
Tested U-value and SHGC for Study 2.

| | Insulation (U-Value) [W/m$^2$·K (Btu/h·ft$^2$·F)] | Solar Gain (SHGC) |
|---|---|---|
| Baseline | No Change | No Change |
| Upgrade 0 | $\leq$ 2.56 (0.45) | No Change |
| Upgrade 1 | $\leq$ 2.56 (0.45) | All windows increased to $\geq$ 0.33 |
| Upgrade 2 | $\leq$ 2.56 (0.45) | Southern windows increased to $\geq$ 0.7; other windows increased to $\geq$ 0.33 |
| Upgrade 3 | Upgrade to Code $\leq$ 2.56 (0.45) | All windows increased ($\geq$ 0.7) |

Simulation Setup for Study 2A and 2B (all with upgraded climate and emission files described in 2.1, and windows with lower U-values and higher SHGC than those specified in each upgrades remain unchanged):

1. Baseline: Using ResStock models with no changes to either insulation (U-value) or solar gain (SHGC) properties.
2. Upgrade 0: Involves upgrading the U-value to 2.56 $W/m^2 \cdot K$ (0.45 $Btu/h \cdot ft^2 \cdot F$),[8] without altering SHGC.
3. Upgrade 1: All windows are upgraded to the U-value of $\leq$ 2.56 $W/m^2 \cdot K$ (0.45 $Btu/h \cdot ft^2 \cdot F$) and a SHGC $\geq$ 0.33.[9]
4. Upgrade 2: Maintains the U-value upgrade to 2.56 $W/m^2 \cdot K$ (0.45 $Btu/h \cdot ft^2 \cdot F$) for all windows and increases the SHGC of only the south-facing windows to 0.7 (same as Study 1), while upgrading the SHGC of other windows to 0.33.
5. Upgrade 3: Upgrades all windows to the U-value $\leq$ 2.56 $W/m^2 \cdot K$ (0.45 $Btu/h \cdot ft^2 \cdot F$) and increases the SHGC of all windows to 0.7.

## 3. Results

### *3.1. Study 1: Impact of high SHGC for South-Facing windows*

#### *3.1.1. Current saving potential – Study 1A*

After conducting simulations on 2,176 instances using the ResStock models representing the 2018 (baseline) residential building stock of Chicago, we investigated the impact of raising the SHGC for south-facing windows to 0.7. Comparing the baseline cases with the cases of increased SHGC, we observe a shift in consumption: the average consumed heating energy decreased from 141.0 $kWh/m^2$ (44.7 $kBtu/ft^2$) to 139.7 $kWh/m^2$ (44.3 $kBtu/ft^2$), while the average consumed cooling energy increased from 47.0 $kWh/m^2$ (14.9 $kBtu/ft^2$) to 47.9 $kWh/m^2$ (15.2 $kBtu/ft^2$). This results in an average reduction of 0.4 $kWh/m^2$ (0.1 $kBtu/ft^2$) in combined energy consumption.

As shown in Fig. 3, the results demonstrate a noticeable reduction in annual emission for heating and cooling with an average of 0.37 kg $CO_2e/m^2$ (0.7 %). The highest saving simulated is 3.86 kg $CO_2e/m^2$ (4.2 %). This finding supports the hypothesis that high SHGC in south-facing windows can contribute positively to reducing emissions in cold climates similar to that of Chicago.

The building variables with the largest impact on carbon savings with upgraded windows were found using Ordinary Least Squares (OLS) linear regression model [42] and Sobol sensitivity analysis [43,44].[10] Notably, buildings with larger window areas on the southern façade exhibited more significant benefits. This outcome is expected, as the modifications were applied exclusively to the southern windows. Larger windows on this façade enable greater solar heat gain during the winter, which helps compensate for the increased cooling demand in the summer. Moreover, buildings with initially low SHGC values experienced significant enhancements in energy performance following the adjustment, suggesting that larger increases in SHGC can result in more substantial savings. After applying filters to focus on upgrading only buildings with southern window area of 2 $m^2$ or more and an SHGC of 0.49 or lower, we observed a notable higher average emission savings, namely, 0.63 kg $CO_2e/m^2$ (1.3 %).

As previously mentioned, increasing the SHGC can create a trade-off between heating and cooling energy use—increasing cooling requirements during the summer while reducing heating requirements in the winter. Consequently, although most cases demonstrated emission reductions from this change, certain buildings experienced adverse effects. Out of the 2,176 cases simulated, only 12 (or 0.6 %) experienced an increase in heating and cooling emissions after raising the SHGC of south-facing windows to 0.7. All 12 cases are apartment units, meaning they are adjacent to other conditioned spaces. As a result, these units are more likely to require cooling rather than heating due to shared thermal boundaries. Most of these cases had a high window-to-wall ratio on the south façade, which resulted in excessive heat gain during the summer. In these 12 cases, this increase in heat gain led to a rise in cooling demands that surpassed the reductions in heating load.

#### *3.1.2. Future saving potential with heat pump upgrades – Study 1B*

Similar to the baseline cases, we simulated the same 2,176 instances using the ResStock heat pump upgraded models and modified the south-facing window SHGC to 0.7. The result shows an overall average heating and cooling energy consumption reduction of 0.19 $kWh/m^2$ (0.06 $kBtu/ft^2$), less than the energy savings in 3.1.1 (buildings without heat pump upgrades). Fig. 4 shows that, because of the SHGC upgrade, there was a notable decrease in annual heating and cooling emissions, averaging 0.15 kg $CO_2e/m^2$ (0.3 %). The highest saving simulated is 1.94 kg $CO_2e/m^2$ (1.9 %). Similar to the findings in section 3.1.1 without heat pump upgrades, with the results from OLS linear regression and Sobol sensitivity analysis,[11] buildings with larger southern window areas show higher potential for savings. A larger change in SHGC from lower SHGC starting values also results in increased emissions savings. In addition, buildings with smaller window areas on the east façade receive greater savings. Focusing on buildings with a southern window area of 2 $m^2$ or more, a baseline SHGC of 0.49 or lower, and an eastern window area of 10 $m^2$ or less, the average emission savings was 0.32 kg $CO_2e/m^2$ (0.7 %).

In the 2,176 heat pump cases, the number of buildings experiencing negative impacts from increasing window SHGC rose from 12 to 71 cases, which is still only 3 % of residences. This increase is likely due to improved heating efficiency, which reduces the potential for heating savings and thus fails (in a small percentage of cases) to compensate for the increased cooling energy consumption.

### *3.2. Study 2: Evaluating combined effects of U-value and SHGC upgrades*

#### *3.2.1. Current saving potential – Study 2A*

In Study 2, we assessed various combinations of U-value and SHGC modifications for 4,826 buildings, since both properties can be selected when upgrading to new windows. The section below (including Table 2 and Fig. 5) presents the results as follows.

1. Upgrade 0 (U-value to code, no SHGC change) exhibited improvements in energy efficiency, as shown by a 3.76 % reduction in annual heating and cooling emissions. This underscores the impact of U-value improvements when implemented without adjustments to SHGC. There is only one case with negative impact,[12] reinforcing the notion that U-value improvements are broadly effective in cold-climate retrofits.

[8] This U-value of 2.56 $W/m^2 \cdot K$ (0.45 $Btu/h \cdot ft^2 \cdot F$) was chosen as point of comparison. It is the 2022 version of ASHRAE 90.1 prescriptive code requirement for Chicago for residential buildings higher than three floors [6].

[9] The SHGC of 0.33 was chosen as a point of comparison. It is the upper limit set by the 2022 version of ASHRAE 90.1 for the Chicago climate zone (ASHRAE Climate Zone 5) for residential buildings higher than three floors [6].

[10] Details shown in Appendix B.

[11] Detailed linear regression and sensitivity analysis shown in Appendix B.

[12] This negative case involves an apartment unit with just one window on the south-facing facade and all other walls adjoining other conditioned spaces. Consequently, the improvement in U-value led to a larger increase in cooling load compared to the modest savings in heating load. This situation resulted in the only negative outcome among over four thousand simulated buildings.

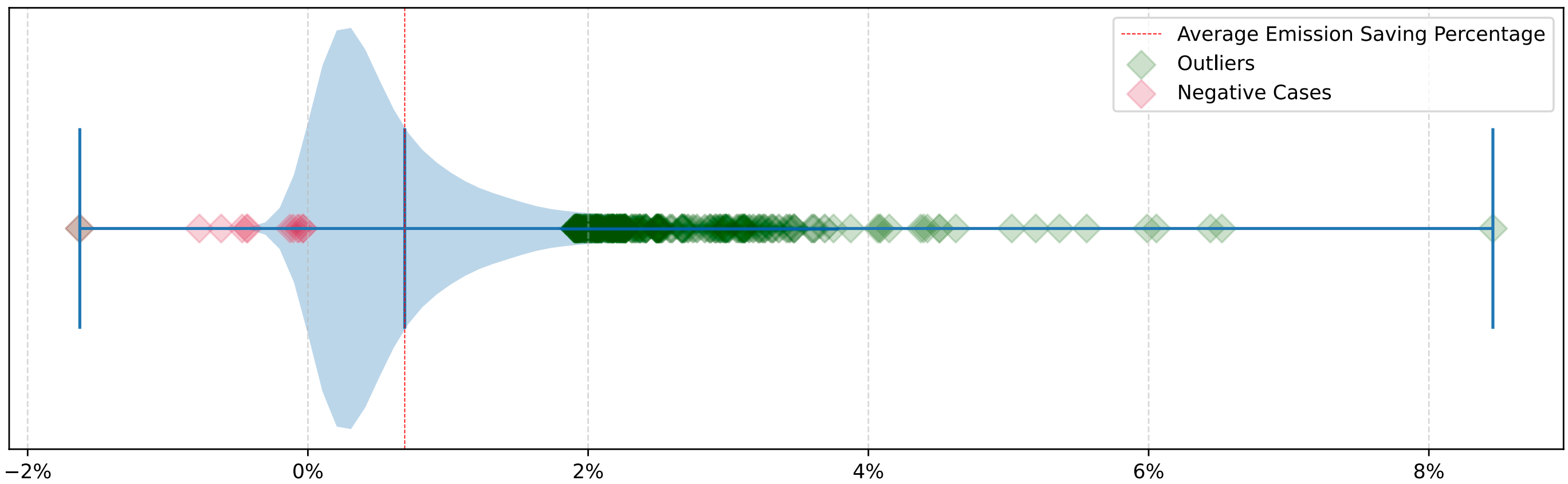

**Fig. 3.** Distribution of emission saving from increasing south window SHGC to 0.7 for the 2018 residential building stock of Chicago.

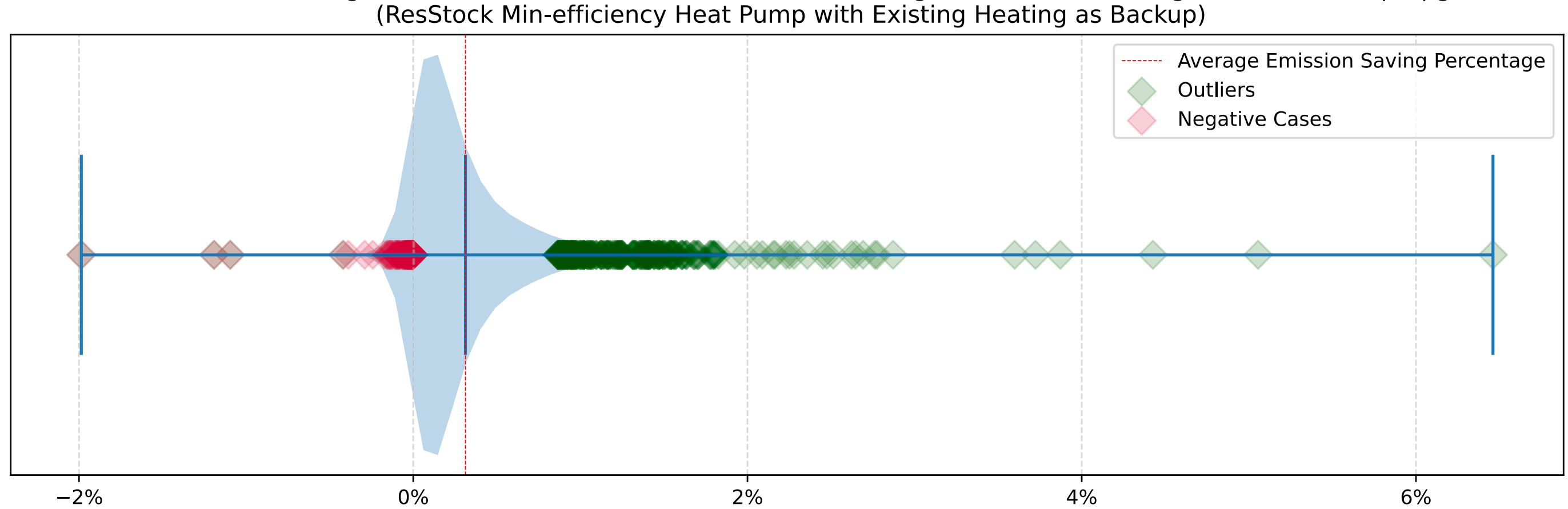

**Fig. 4.** Distribution of emission saving from increasing south window SHGC to 0.7 for heat-pump-upgraded Chicago residential building stock.

**Table 2**
Emission saving results for the 2018 residential building stock of Chicago with window upgrades in Study 2.

| | **Insulating Characteristics** U-Value [W/m$^2$·K (Btu/h·ft$^2$·F)] | **Solar Heat Gain [SHGC]** | **Average Emissions Reduction per Building Floor Area** [kg $CO_2$e/ m$^2$] | **% Savings** |
|---|---|---|---|---|
| Upgrade 0 | ≤ 2.56 (0.45) | No Change | 2.10 | **3.76 %** |
| Upgrade 1 | ≤ 2.56 (0.45) | ≥ 0.33 | 2.10 | **3.77 %** |
| Upgrade 2 | ≤ 2.56 (0.45) | South windows ≥ 0.7; other windows ≥ 0.33 | 2.26 | **4.07 %** |
| Upgrade 3 | ≤ 2.56 (0.45) | ≥ 0.7 | 2.54 | **4.59 %** |

2. Upgrade 1 (All windows to 0.33 SHGC and improved U-value) led to moderate improvements, achieving a 3.77 % emissions reduction. This outcome indicates that while bringing both U-values and SHGC to code-inspired standards provides some benefits in balancing winter heating and summer cooling demands, the improvements are not significantly better than improving U-value alone. The same negative case in Upgrade 0 is also present in Upgrade 1.
3. Upgrade 2 (improved U-value, high [0.7] SHGC in south-facing windows, moderate [0.33] SHGC in other windows) resulted in a slightly higher heating and cooling emissions reduction of 4.07 % compared to the baseline, highlighting the efficiency of targeting south-facing windows with higher SHGC values. This finding supports our hypothesis that combining better U-value with high SHGC can result in more emission savings. While the average emission savings for Upgrade 2 are greater than those for Upgrades 0 and 1, it also increased the number of negative cases from 1 to 9, still < 0.2 % of cases.
4. Upgrade 3 (U-value to code, all windows to high SHGC) resulted in the highest emissions reduction of 4.59 %, at the cost of 68 negative cases (1.4 %).

Notably, some of the simulated cases lacked cooling systems, reflecting the current state of certain residential buildings in Chicago.[13] Thus, with high SHGC, these buildings benefited from reduced heating loads in winter without any increase in cooling demands, as they originally had no cooling systems.

[13] The results presented in section 3.2.2 will detail the cases involving heat pump upgrades, ensuring that every building is equipped with a cooling system. This setup will more accurately capture the trade-offs between heating and cooling demands.

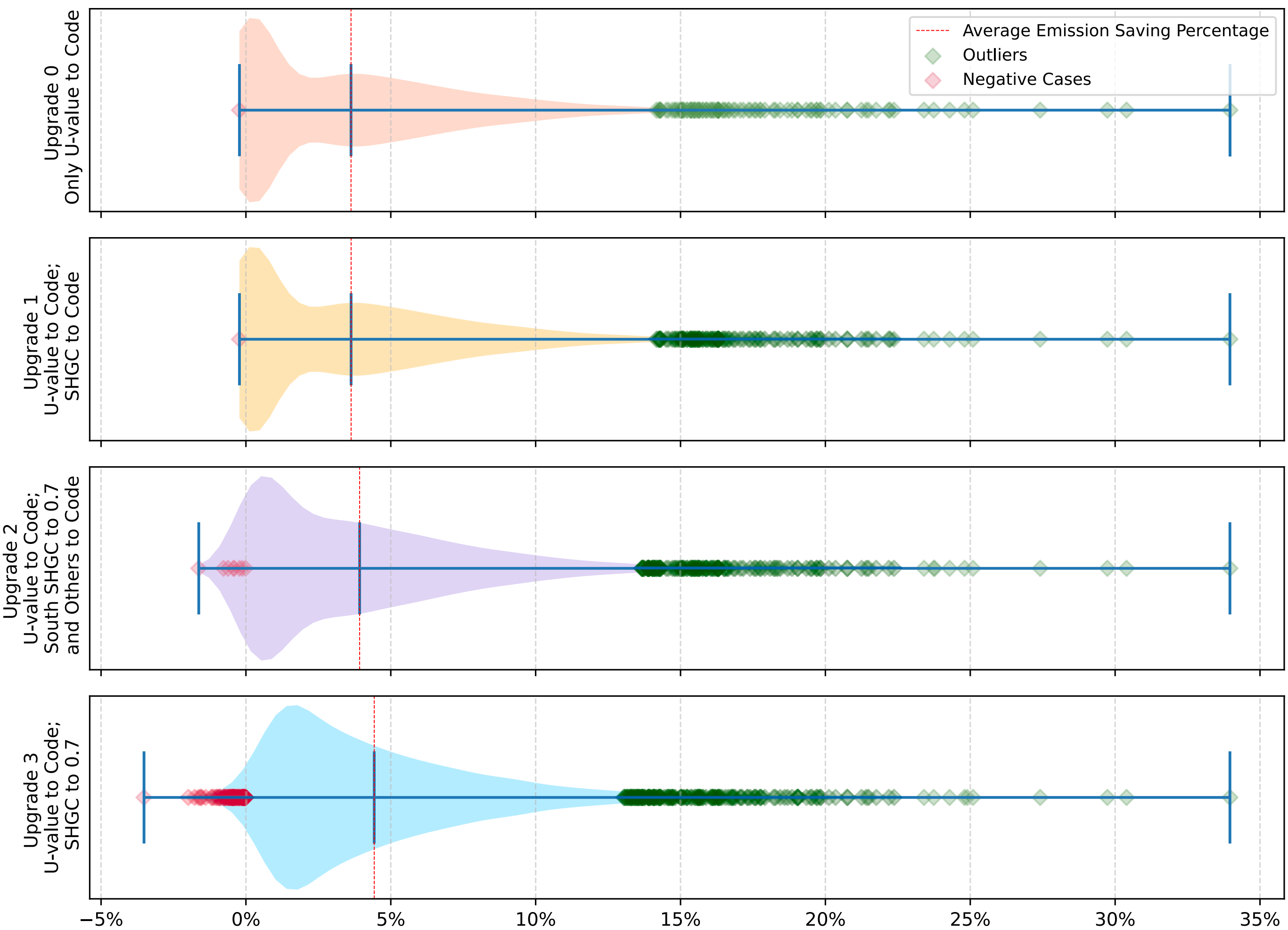

**Fig. 5.** Distribution of emission savings of the 2018 Chicago residential building stock for window upgrades with U-value upgrade and various SHGC.

Further linear regression and sensitivity analysis[14] on Upgrade 3 reveals that buildings with higher U-value for the windows, higher air leakage value for the entire building, and lower SHGC achieve greater emission savings. This highlights that significant changes in U-value and SHGC have a more pronounced impact. In essence, leaky buildings with poorly insulated windows and low SHGC stand to benefit the most from upgrading to windows with lower U-values and higher SHGC, especially in the climate of Chicago. Since U-value and SHGC are correlated in the ResStock model, we prioritized filtering for higher U-values, as it was identified as the most significant variable in the sensitivity analysis. By selecting buildings only with U-values above 2.56 W/m$^2$·K (requiring upgrades to meet ASHRAE 90.1 standards) and building air leakage values exceeding 5 Air Changes per Hour (ACH), the average emission savings achieved by Upgrade 3 (U-value to code, all windows to high SHGC) increased to 2.99 kg CO2e/m$^2$ (5.2 %).

### *3.2.2. Future saving potential with heat pump upgrades – Study 2B*

Next, repeating the analysis but including the upgrades to building heat pumps, Table 3 presents the numerical results, and Fig. 6 illustrates the distribution of emission savings for the various upgrades discussed in the study:

**Table 3**
Emission saving results for cases with heat pump (and window) upgrades in Study 2.

| | **Insulating Characteristics** U-Value [W/m$^2$·K (Btu/h·ft$^2$·F)] | **Solar Heat Gain [SHGC]** | **Average Emissions Reduction per Building Floor Area** [kg CO$_2$e/ m$^2$] | **% Savings** |
|---|---|---|---|---|
| Upgrade 0 | $\leq$ 2.56 (0.45) | No Change | 1.12 | **2.20 %** |
| Upgrade 1 | $\leq$ 2.56 (0.45) | $\geq$ 0.33 | 1.12 | **2.20 %** |
| Upgrade 2 | $\leq$ 2.56 (0.45) | South windows $\geq$ 0.7; other windows $\geq$ 0.33 | 1.2 | **2.40 %** |
| Upgrade 3 | $\leq$ 2.56 (0.45) | $\geq$ 0.7 | 1.27 | **2.50 %** |

1. Upgrade 0 (U-value to code, no SHGC change) exhibited a 2.20 % reduction in emissions. There is only one case with negative impact

[14] Details in Appendix B.

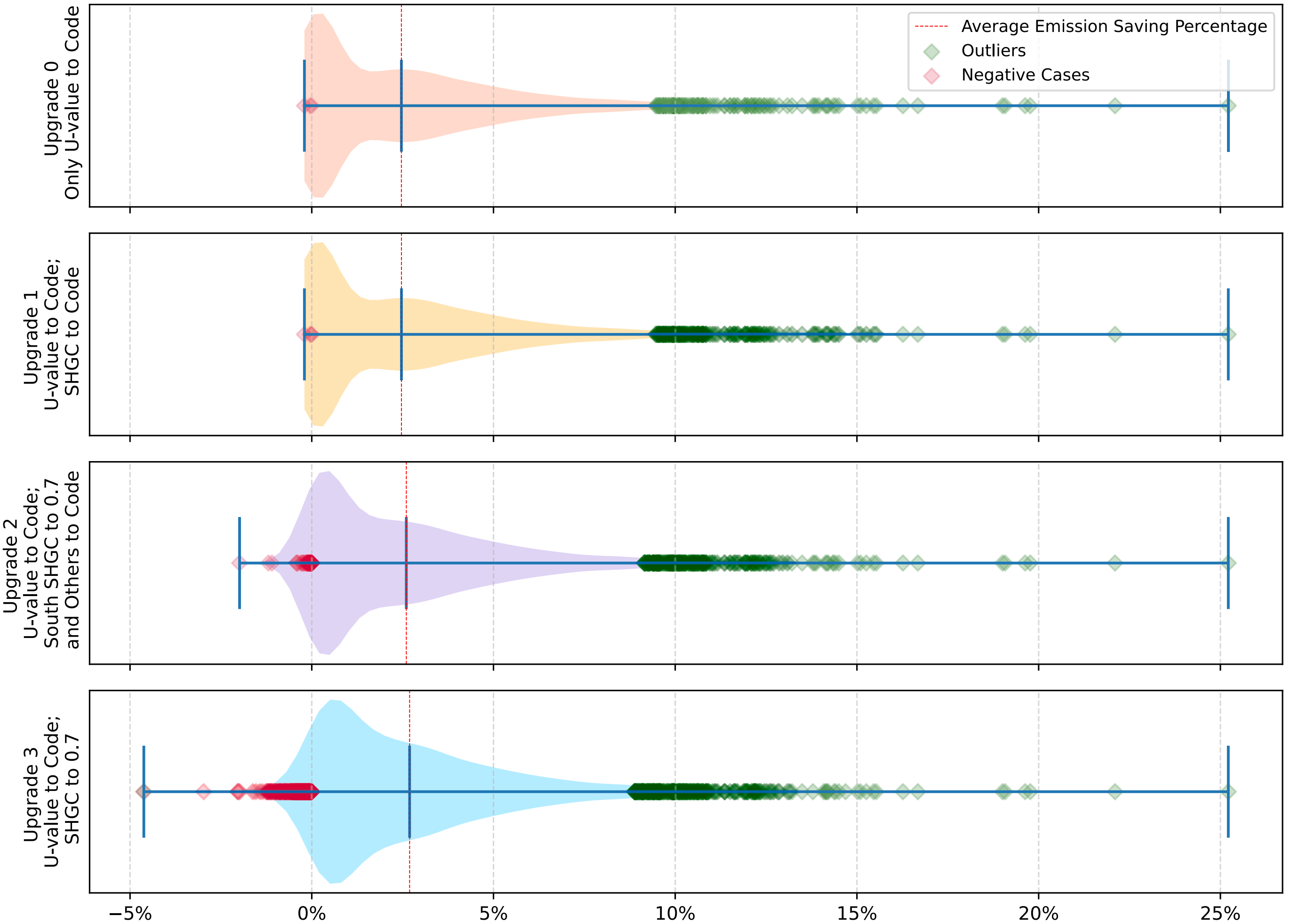

**Fig. 6.** Distribution of emission savings of heat-pump-upgraded Chicago residential buildings for window upgrades with U-value upgrades and various SHGC.

(detailed description in 3.2.1), suggesting that window insulation improvement alone can benefit nearly all scenarios.

2. Upgrade 1 (All windows to code SHGC and U-value) led to moderate improvements with the same negative case, achieving the same 2.20 % emissions reduction as Upgrade 0, indicating the improvements are not significantly better than improving U-value alone. Since the average SHGC before modification was 0.60—most of the buildings already had an SHGC higher than 0.33—the impact for Upgrade 1 is relatively minimal.
3. Upgrade 2 (High SHGC in south-facing windows, standard in others) resulted in a slightly higher emissions reduction of 2.40 %. Fig. 6 shows that although the average emission savings for Upgrade 2 are greater, it also includes 29 negative cases compared to Upgrade 0 and 1.
4. Upgrade 3 (All windows high SHGC) resulted in the highest emissions reduction of 2.50 %. However, the number of negative cases also increased to 283 (5.9 %). Here, increasing SHGC for all windows should be approached selectively to mitigate potential negative impacts. Two factors may explain the variation in emission reductions between the current scenario (3.2.1) and the heat-pump-upgraded scenario (3.2.2): firstly, the heat pump upgrade shifts heating from dirtier energy sources—such as natural gas—to electricity, thereby reducing initial heating emissions before the SHGC changes; secondly, because all buildings are equipped with cooling systems following the heat pump upgrade, the trade-offs from increased cooling demands become more significant.

Similar to the findings in 3.2.1 using the same regression and sensitivity analysis,[15] the leading factors for more emission savings in Upgrade 3 are higher U-value for the windows, higher building air leakage value, and lower SHGC. Filtering buildings with window U-values above 2.56 $W/m^2{\cdot}K$ and building air leakage values exceeding 5 ACH, the average emission savings achieved by Upgrade 3 (U-value to code, all windows to high SHGC) increased to 1.67 kg $CO2e/m^2$ (3.5 %).

## 4. Discussion

### *4.1. Result Interpretation and practical Implications*

The findings from the investigations in Studies 1 and 2 highlight the vital role of window design optimization in enhancing energy efficiency and reducing emissions in residential buildings. In Study 1, increasing the SHGC of south-facing windows to 0.7 (significantly higher than the values appearing in other references [14,17], as noted in Section 1) demonstrated a net reduction in total energy consumption and long-run marginal carbon emissions. These energy and emissions savings were predominantly due to the increased solar gains during Chicago's colder months, which decreased heating energy use more than offsetting the increase in cooling energy use. Notably, buildings with specific architectural features such as lower initial SHGC showed the most substantial benefits.

[15] Regression model and sensitivity analysis details in Appendix B.

While both studies considered future climate scenarios and long-run marginal emissions, Study 2 extended this analysis by exploring combinations of U-value and SHGC modifications. The results revealed that while U-value improvements alone provided benefits, strategic combinations of high SHGC in south-facing windows significantly enhanced overall building performance. Particularly, balancing between reducing winter heating loads and managing summer cooling demands can lead to the most favorable emission reductions.

The results support both of the initial hypotheses. The first hypothesis, which suggested that higher SHGC in windows than typically recommended could reduce overall emissions from heating and cooling, was validated. Importantly, while initially the hypothesis focused on south-facing windows, findings from Study 2 demonstrated that upgrading windows to 0.7 SHGC in *all* orientations led to more significant emission reductions. Put another way, 0.33 SHGC was substantially too low for maximizing savings, even for the non-south-facing windows.

The second hypothesis was also confirmed, showing that retrofit windows combining higher SHGC with improved insulation values perform better in reducing long-run marginal emissions than those with improved insulation alone. This finding has practical importance, because, while retrofitting windows solely for SHGC improvements may not be cost-effective, integrating SHGC optimization during window retrofits for other benefits (which could also include aesthetics or airtightness) can enhance overall building performance. As noted in Section 1, past studies [14,17] have paired improved window U-values with *decreased* SHGC values. While these were reasonable test parameters for those studies, readers should not misinterpret past findings to infer that lower SHGC values for window retrofits in existing residential buildings in cold climates are optimal for energy and emissions savings.

The results of this study confirm that better insulating performance (i.e. lower U-Value) is an undisputedly important strategy for reducing emissions in cold climate retrofits. Meanwhile, a review of the product libraries of several window manufacturers revealed that lower U-Values (desirable) are often correlated with lower SHGC values (undesirable in our case studies), since adding an additional pane or low-emissivity coating (strategies employed for improving insulation performance) will usually reduce the window's SHGC value. However, importantly, the two properties (U-Value and SHGC) are not 100 % linked, and as discussed in Section 1, products exist from multiple manufacturers that can meet the specifications simulated here. Moreover, emerging technologies like clearer glass, better frames, gas fills, and vacuum-insulated glazing units, may allow for further improvements to U-Value without a significant reduction in SHGC.

As expected, for both Study 1 and 2, greater emission savings were observed in scenarios without heat pump upgrades. This outcome likely occurs for two reasons. First, reductions in heating load have a more substantial impact when the baseline heating system is less efficient and/or the energy source, such as on-site natural gas, is more carbon intensive than an electric heat pump system. Second, as noted, the heat pump upgrade cases ensure that mechanical cooling is present and any increases in cooling demand are accounted for in all cases. Nevertheless, it is important to note that window retrofits, and specifically window retrofits with high SHGC values still showed significant emissions savings in these cold climate cases, even if the homes are upgraded with heat pumps.[16] Together, these results highlight the effectiveness of strategic window upgrades in enhancing building energy efficiency and reducing environmental impact across various scenarios.

### 4.2. Limitations and future work

As noted in Section 2.2, the simulations in this study used future weather files, i.e. warmer conditions than the typical meteorological year weather files usually used for building energy simulation [36]. This is a conservative approach, since the tested SHGC retrofits allow for more passive solar heat gain. However, choosing an even more extreme warming scenario would likely reduce savings.

In an effort to capture the daily and seasonal differences in the impacts of electricity, this study converted simulated energy consumption to carbon emissions, specifically using NREL's Cambium long-run marginal emission factors, which are emission-time-sensitive, and based on an analysis of long-term emission impacts [20]. This is one estimation of carbon emissions, and future work could expand on the analysis using different conversion methods.

It is important to note that this study focused on window replacements and did not analyze whether the existing maximums for SHGC in standards like ASHRAE [6,7] or codes like IECC [8] are ideal for new buildings or major building renovations. Major renovations targeted by these codes would also require significant changes to other building elements like wall and roof insulation as well as HVAC efficiency, effectively resulting in a substantially different building. In contrast, this study focused on limited retrofits to existing residential buildings.

Window SHGC influences indoor temperatures, affecting both heat and cold resilience in buildings. In Chicago, where many buildings lack cooling systems, an increase in SHGC can lead to heightened indoor heat, particularly during warmer months. This underscores the importance of considering heat resilience when selecting SHGC values, especially for climate adaptation in urban settings. Some studies have explored heat resilience using Chicago's ResStock models [45,46], aiming to balance energy efficiency with comfort and safety.

Additionally, this study did not account for human behavior, such as how occupants manage their blinds. While we assumed blinds remain open in winter to maximize passive solar heat gain, occupants may close them to reduce glare from the low-angle winter sun, which would reduce this heat gain. Such behavioral variations are beyond the scope of our research.

This study demonstrates that higher SHGC can yield energy and emission savings, but it does not determine the optimal SHGC for each orientation or urban context, suggesting that project teams could achieve further refinement. Future research could extend to parametric testing of various window properties and incorporate investigations into additional sun shading options. By simulating a diverse range of window characteristics and shading techniques tailored to specific building orientations and urban contexts, project teams could enhance the precision of their energy efficiency and emissions reduction strategies, optimizing solar gains and mitigating potential overheating.

## 5. Conclusion

In conclusion, this research demonstrates that retrofitting windows, and in particular increasing their SHGC, can substantially improve the operational carbon emissions of residential buildings in a cold city. The methods demonstrate the use of future climate projections and long-run marginal emissions in building design decision-making. As the building sector and electricity grid continue to evolve, such integrated approaches will be important in driving forward the sustainability agenda, particularly in response to the escalating challenges posed by climate change.

**Author Contributions**

Yiwei Lyu conducted literature review for the project, contributed to the concept and experiment development, analyzed the data, visualized the results, wrote and edited the paper. Jialiang Xiang contributed to the concept and experiment development, managed the simulations, parsed the data, helped with data analysis, and edited the paper. Holly Samuelson led this research project, contributed to the concept and experiment development, edited the paper, and secured funding.

[16] In fact, window upgrades have the added benefit of reducing peak heating loads, which could reduce the cost of an eventual heat pump retrofit.

### CRediT authorship contribution statement

**Yiwei Lyu:** Writing – review & editing, Writing – original draft, Visualization, Software, Methodology, Data curation, Conceptualization. **Jialiang Xiang:** Visualization, Software, Methodology, Investigation, Data curation, Conceptualization. **Holly Samuelson:** Writing – review & editing, Supervision, Project administration, Methodology, Funding acquisition, Conceptualization.

### Declaration of competing interest

The authors declare that they have no known competing financial interests or personal relationships that could have appeared to influence the work reported in this paper.

### Acknowledgements

This work was supported by the Google + Harvard, Climate + Data project; the Harvard Center for Green Buildings and Cities; and the Salata Institute for Climate and Sustainability. We are grateful for their support. We thank Sheng Liu for his help finding international resources. Statistical support was provided by data science specialist Dan Yuan, at the Institute for Quantitative Social Science, Harvard University.

## Appendix A

**Window Products with Low U-value and High SHGC.**

To identify window products with high SHGC and low U-values, we used the list provided by Window Digest for the top 21 window glass manufacturers [47]. Because companies including One day glass, Valley Glass Company, East Bay Glass, Restoration Window Glass, American Window & Glass, Northern Comfort Windows & Doors, Panes Window Manufacturing, AGW Glass, INTIGRAL, and Port Window Glass, Dallas Flat Glass do not provide easy access to their thermal performance, we listed 10 products from each of the rest of the companies. All windows are selected to achieve U-values below 2.5 W/m$^2$·K (0.45 Btu/h·ft$^2$·F), ensuring compliance with ASHRAE 90.1 standards for Climate Zone 5 [6]. Additionally, many of these windows meet the more stringent Northern Climate Zone Energy Star requirements [10], with U-values as low as 1.5 W/m$^2$·K (0.26 Btu/h·ft$^2$·F).

| Window Manufacturer | Product Name | U-value [W/m$^2$·K (Btu/h·ft$^2$·F)] | SHGC |
|---|---|---|---|
| Vitro [48] | SUNGATE 400 (2) STARPHIRE + STARPHIRE – Winter Argon | 1.6 (0.28) | 0.68 |
| Guardian Glass [49] | Guardian Clear Glass (North America) Glass, 1/8″ (3 mm)– 10 % Air, 90 % Argon 12.7 mm – Guardian Clear Glass (North America) Glass, 1/8″ (3 mm) (3-ClimaGuard® 80/71 (North America)) | 1.5 (0.26) | 0.71 |
| Cardinal Glass Industries [18] | Clear / LoE-180® (#3) | 1.5 (0.26) | 0.68 |
| AIS Glass [50] | Clear Essence Plus – 6 mm (Low-E Glass) / 12 mm (Air Gap) / 6 mm (Clear Glass) | 1.8 (0.32) | 0.61 |
| Milgard [51] | Radius Picture Window – 7/8″ – Cardinal 180/4th Surface Foam Argon | 1.6 (0.28) | 0.56 |
| Pilkington [19] | Pilkington Energy Advantage™ Low-e (coating on #2 surface) outer lite and Pilkington Energy Advantage™ Low-e (coating on #4 surface) inner lite9 with argon fill | 1.4 (0.23) | 0.66 |
| United Plate Glass Company [52] | Energy Advantage / Clear IGU | 1.9 (0.33) | 0.62 |
| Pella [53] | Fixed Dual-Pane Glazing – Wood Exterior – 11/16″ – Clear IG | 2.5 (0.44) | 0.62 |
| Hartung Glass [54] | Bird1st UV with NU 78/65 (#5) | 1.7 (0.30) | 0.59 |
| Press Glass [55] | 4/10Kr/4T 1,1 | 1.0 (0.18) | 0.66 |

## Appendix B

**Linear Regression Model and Sensitivity Analysis Results.**

**Study 1A** — increasing south window SHGC to 0.7 for the 2018 residential building stock of Chicago:

Initial OLS linear regression model results (includes every relevant variable):

R-squared is 0.388.

| Variable | Coef | Std Err | t | P>\|t\| | [0.025 | 0.975] |
|---|---|---|---|---|---|---|
| const | 4.89E + 09 | 1.4E + 11 | 0.035 | 0.972 | −2.7E + 11 | 2.8E + 11 |
| in.tenure_Metadata | −0.0042 | 0.003 | −1.293 | 0.196 | −0.01 | 0.002 |
| in.usage_level_Metadata | −0.0055 | 0.004 | −1.541 | 0.123 | −0.013 | 0.002 |
| conditioned_floor_area | −0.0613 | 0.014 | −4.391 | 0 | −0.089 | −0.034 |
| air_leakage_value | 0.0896 | 0.014 | 6.525 | 0 | 0.063 | 0.117 |
| heating_primary_efficiency_value | −0.1383 | 0.044 | −3.158 | 0.002 | −0.224 | −0.052 |
| cooling_primary_efficiency_value | 0.0601 | 0.014 | 4.355 | 0 | 0.033 | 0.087 |
| year_normalized | −0.0087 | 0.005 | −1.742 | 0.082 | −0.019 | 0.001 |
| income_normalized | −0.0023 | 0.004 | −0.529 | 0.597 | −0.011 | 0.006 |
| roof_assembly_effective_rvalue | 0.0052 | 0.005 | 1.028 | 0.304 | −0.005 | 0.015 |
| average_floor_rvalue | −0.0147 | 0.008 | −1.789 | 0.074 | −0.031 | 0.001 |
| shgc_before | −0.1769 | 0.014 | −13.044 | 0 | −0.204 | −0.15 |

(continued on next page)

(*continued*)

| Variable | Coef | Std Err | t | P>\|t\| | [0.025 | 0.975] |
|---|---|---|---|---|---|---|
| u_factor | 0.0126 | 0.008 | 1.557 | 0.12 | −0.003 | 0.028 |
| projection_factor_east | −0.0033 | 0.007 | −0.483 | 0.629 | −0.017 | 0.01 |
| projection_factor_north | −0.0071 | 0.006 | −1.13 | 0.258 | −0.019 | 0.005 |
| projection_factor_south | −0.0188 | 0.006 | −2.999 | 0.003 | −0.031 | −0.007 |
| projection_factor_west | 0.0076 | 0.007 | 1.134 | 0.257 | −0.006 | 0.021 |
| total_area_east | −0.1052 | 0.035 | −2.967 | 0.003 | −0.175 | −0.036 |
| total_area_north | 0.1108 | 0.046 | 2.389 | 0.017 | 0.02 | 0.202 |
| total_area_south | 0.1338 | 0.045 | 2.945 | 0.003 | 0.045 | 0.223 |
| total_area_west | −0.1189 | 0.035 | −3.433 | 0.001 | −0.187 | −0.051 |
| window_to_wall_ratio_east | 0.0312 | 0.012 | 2.681 | 0.007 | 0.008 | 0.054 |
| window_to_wall_ratio_north | 0.0069 | 0.014 | 0.482 | 0.63 | −0.021 | 0.035 |
| window_to_wall_ratio_south | 0.055 | 0.011 | 4.924 | 0 | 0.033 | 0.077 |
| window_to_wall_ratio_west | 0.0202 | 0.012 | 1.758 | 0.079 | −0.002 | 0.043 |
| heating_primary_fuel_electricity | −2.5E + 09 | 7.05E + 10 | −0.035 | 0.972 | −1.4E + 11 | 1.36E + 11 |
| heating_primary_fuel_natural gas | −2.5E + 09 | 7.05E + 10 | −0.035 | 0.972 | −1.4E + 11 | 1.36E + 11 |
| heating_primary_fuel_propane | −2.5E + 09 | 7.05E + 10 | −0.035 | 0.972 | −1.4E + 11 | 1.36E + 11 |
| residential_facility_type_apartment unit | −1.7E + 09 | 4.95E + 10 | −0.035 | 0.972 | −9.9E + 10 | 9.53E + 10 |
| residential_facility_type_single-family attached | −1.7E + 09 | 4.95E + 10 | −0.035 | 0.972 | −9.9E + 10 | 9.53E + 10 |
| residential_facility_type_single-family detached | −1.7E + 09 | 4.95E + 10 | −0.035 | 0.972 | −9.9E + 10 | 9.53E + 10 |
| heating_primary_type_Boiler | −5.5E + 08 | 1.58E + 10 | −0.035 | 0.972 | −3.2E + 10 | 3.04E + 10 |
| heating_primary_type_ElectricResistance | −5.5E + 08 | 1.58E + 10 | −0.035 | 0.972 | −3.2E + 10 | 3.04E + 10 |
| heating_primary_type_Furnace | −5.5E + 08 | 1.58E + 10 | −0.035 | 0.972 | −3.2E + 10 | 3.04E + 10 |
| heating_primary_type_HeatPump | −3.5E + 08 | 1E + 10 | −0.035 | 0.972 | −2E + 10 | 1.93E + 10 |
| heating_primary_type_WallFurnace | −5.5E + 08 | 1.58E + 10 | −0.035 | 0.972 | −3.2E + 10 | 3.04E + 10 |
| cooling_primary_type_0 | −1.6E + 08 | 4.56E + 09 | −0.035 | 0.972 | −9.1E + 09 | 8.78E + 09 |
| cooling_primary_type_HeatPump | −3.6E + 08 | 1.03E + 10 | −0.035 | 0.972 | −2.1E + 10 | 1.98E + 10 |
| cooling_primary_type_central air conditioner | −1.6E + 08 | 4.56E + 09 | −0.035 | 0.972 | −9.1E + 09 | 8.78E + 09 |
| cooling_primary_type_mini-split | −1.6E + 08 | 4.56E + 09 | −0.035 | 0.972 | −9.1E + 09 | 8.78E + 09 |
| cooling_primary_type_room air conditioner | −1.6E + 08 | 4.56E + 09 | −0.035 | 0.972 | −9.1E + 09 | 8.78E + 09 |

Refitted OLS model (removing P>|t| less than 0.05):

R-squared is 0.344.

| Variable | Coef | Std Err | t | P>\|t\| | [0.025 | 0.975] |
|---|---|---|---|---|---|---|
| const | 0.1818 | 0.008 | 22.833 | 0 | 0.166 | 0.197 |
| conditioned_floor_area | −0.0945 | 0.013 | −7.072 | 0 | −0.121 | −0.068 |
| air_leakage_value | 0.0931 | 0.014 | 6.778 | 0 | 0.066 | 0.12 |
| heating_primary_efficiency_value | −0.1097 | 0.011 | −9.857 | 0 | −0.132 | −0.088 |
| cooling_primary_efficiency_value | −0.0096 | 0.004 | −2.584 | 0.01 | −0.017 | −0.002 |
| shgc_before | −0.1829 | 0.014 | −13.206 | 0 | −0.21 | −0.156 |
| u_factor | 0.0167 | 0.008 | 2.027 | 0.043 | 0.001 | 0.033 |
| projection_factor_south | −0.0134 | 0.005 | −2.844 | 0.005 | −0.023 | −0.004 |
| total_area_south | 0.1953 | 0.026 | 7.485 | 0 | 0.144 | 0.246 |
| total_area_west | −0.0779 | 0.016 | −4.831 | 0 | −0.11 | −0.046 |
| window_to_wall_ratio_east | 0.0146 | 0.005 | 2.653 | 0.008 | 0.004 | 0.025 |
| window_to_wall_ratio_south | 0.0511 | 0.007 | 7.725 | 0 | 0.038 | 0.064 |

Top 3 coefficients from refitted OLS model (largest absolute value):

total_area_south, shgc_before, const.

Result of Sobol sensitivity analysis:

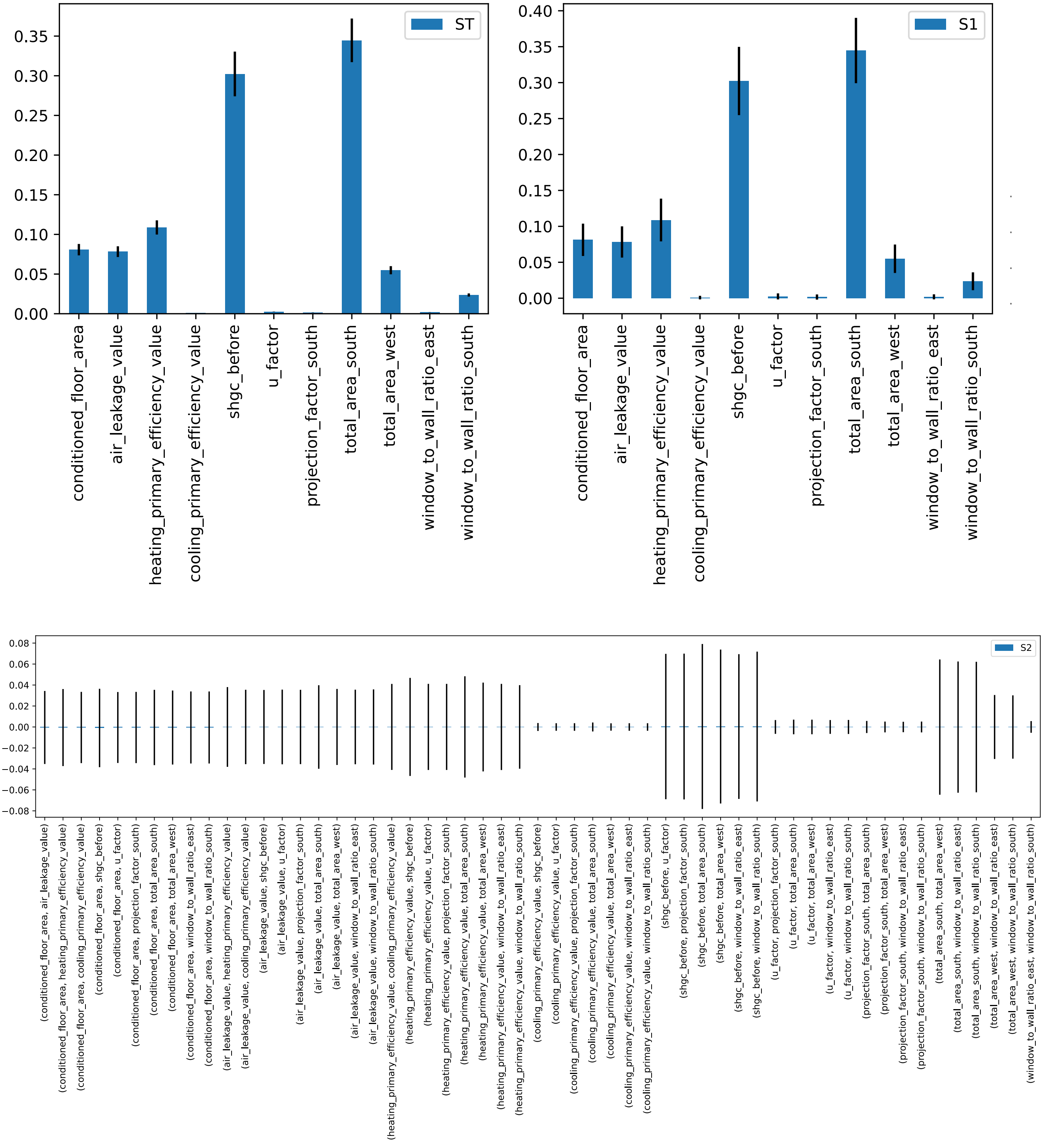

**Fig. 7.** Sobol sensitivity analysis plot of Study 1A: increasing south window SHGC to 0.7 for the 2018 residential building stock of Chicago.

Most influential variables for emission savings are the total window area on the south façade and SHGC before the retrofit.
**Study 1B** — increasing south window SHGC to 0.7 for heat-pump-upgraded Chicago residential building stock:
Initial OLS linear regression model results (includes every relevant variable):
R-squared is 0.556.

| Variable | Coef | Std Err | t | P>\|t\| | [0.025 | 0.975] |
|---|---|---|---|---|---|---|
| const | −5.9E + 09 | 7.02E + 10 | −0.084 | 0.933 | −1.4E + 11 | 1.32E + 11 |
| in.tenure_Metadata | −0.0015 | 0.001 | −1.003 | 0.316 | −0.004 | 0.001 |
| in.usage_level_Metadata | −0.0026 | 0.002 | −1.597 | 0.11 | −0.006 | 0.001 |
| conditioned_floor_area_Before | −0.0435 | 0.008 | −5.435 | 0 | −0.059 | −0.028 |
| air_leakage_value_ACH_Before | 0.0325 | 0.006 | 5.158 | 0 | 0.02 | 0.045 |
| heating_backup_type_Before | 0.0283 | 0.001 | 20.188 | 0 | 0.026 | 0.031 |
| heating_backup_efficiency_value_Before | −0.0016 | 0.004 | −0.441 | 0.659 | −0.009 | 0.006 |
| year_normalized | 0.0038 | 0.002 | 1.672 | 0.095 | −0.001 | 0.008 |
| income_normalized | −0.0025 | 0.002 | −1.218 | 0.223 | −0.006 | 0.002 |
| roof_assembly_effective_rvalue | 0.0009 | 0.002 | 0.382 | 0.703 | −0.004 | 0.005 |
| average_floor_rvalue | 0.0002 | 0.004 | 0.042 | 0.967 | −0.008 | 0.008 |
| shgc_before | −0.1063 | 0.006 | −17.651 | 0 | −0.118 | −0.094 |
| u_factor | 0.0111 | 0.004 | 3.107 | 0.002 | 0.004 | 0.018 |
| projection_factor_east | −0.0054 | 0.003 | −1.552 | 0.121 | −0.012 | 0.001 |
| projection_factor_north | −0.0075 | 0.003 | −2.25 | 0.025 | −0.014 | −0.001 |
| projection_factor_south | −0.0112 | 0.003 | −3.347 | 0.001 | −0.018 | −0.005 |
| projection_factor_west | −0.0049 | 0.003 | −1.436 | 0.151 | −0.012 | 0.002 |
| total_area_east | −0.0782 | 0.016 | −4.932 | 0 | −0.109 | −0.047 |
| total_area_north | 0.0856 | 0.021 | 4.092 | 0 | 0.045 | 0.127 |
| total_area_south | 0.1036 | 0.022 | 4.713 | 0 | 0.06 | 0.147 |
| total_area_west | −0.0712 | 0.016 | −4.464 | 0 | −0.102 | −0.04 |
| window_to_wall_ratio_east | 0.0197 | 0.005 | 3.701 | 0 | 0.009 | 0.03 |
| window_to_wall_ratio_north | 0.0084 | 0.007 | 1.248 | 0.212 | −0.005 | 0.022 |
| window_to_wall_ratio_south | 0.0101 | 0.006 | 1.718 | 0.086 | −0.001 | 0.022 |
| window_to_wall_ratio_west | 0.0161 | 0.005 | 3.004 | 0.003 | 0.006 | 0.027 |
| residential_facility_type_Before_apartment unit | 3.3E + 09 | 3.94E + 10 | 0.084 | 0.933 | −7.4E + 10 | 8.06E + 10 |
| residential_facility_type_Before_single-family attached | 3.3E + 09 | 3.94E + 10 | 0.084 | 0.933 | −7.4E + 10 | 8.06E + 10 |
| residential_facility_type_Before_single-family detached | 3.3E + 09 | 3.94E + 10 | 0.084 | 0.933 | −7.4E + 10 | 8.06E + 10 |
| heating_backup_fuel_Before_electricity | 2.58E + 09 | 3.08E + 10 | 0.084 | 0.933 | −5.8E + 10 | 6.3E + 10 |
| heating_backup_fuel_Before_natural gas | 2.58E + 09 | 3.08E + 10 | 0.084 | 0.933 | −5.8E + 10 | 6.3E + 10 |
| heating_backup_fuel_Before_propane | 2.58E + 09 | 3.08E + 10 | 0.084 | 0.933 | −5.8E + 10 | 6.3E + 10 |

Refitted OLS model (removing P>|t| less than 0.05):

R-squared is 0.511.

| Variable | Coef | Std Err | t | P>\|t\| | [0.025 | 0.975] |
|---|---|---|---|---|---|---|
| const | 0.0696 | 0.003 | 20.123 | 0 | 0.063 | 0.076 |
| conditioned_floor_area_Before | −0.0547 | 0.008 | −7.163 | 0 | −0.07 | −0.04 |
| air_leakage_value_ACH_Before | 0.03 | 0.006 | 4.768 | 0 | 0.018 | 0.042 |
| heating_backup_type_Before | 0.0263 | 0.001 | 22.12 | 0 | 0.024 | 0.029 |
| shgc_before | −0.107 | 0.006 | −17.121 | 0 | −0.119 | −0.095 |
| u_factor | 0.0114 | 0.004 | 3.084 | 0.002 | 0.004 | 0.019 |
| total_area_east | −0.0988 | 0.016 | −6.246 | 0 | −0.13 | −0.068 |
| total_area_north | 0.0822 | 0.009 | 8.712 | 0 | 0.064 | 0.101 |
| total_area_south | 0.138 | 0.012 | 11.982 | 0 | 0.115 | 0.161 |
| total_area_west | −0.0618 | 0.016 | −3.863 | 0 | −0.093 | −0.03 |
| window_to_wall_ratio_east | 0.0292 | 0.005 | 6.081 | 0 | 0.02 | 0.039 |
| window_to_wall_ratio_west | 0.0146 | 0.005 | 2.997 | 0.003 | 0.005 | 0.024 |

Top 3 coefficients from refitted OLS model (largest absolute value):

total_area_south, shgc_before, total_area_east.

Result of Sobol sensitivity analysis:

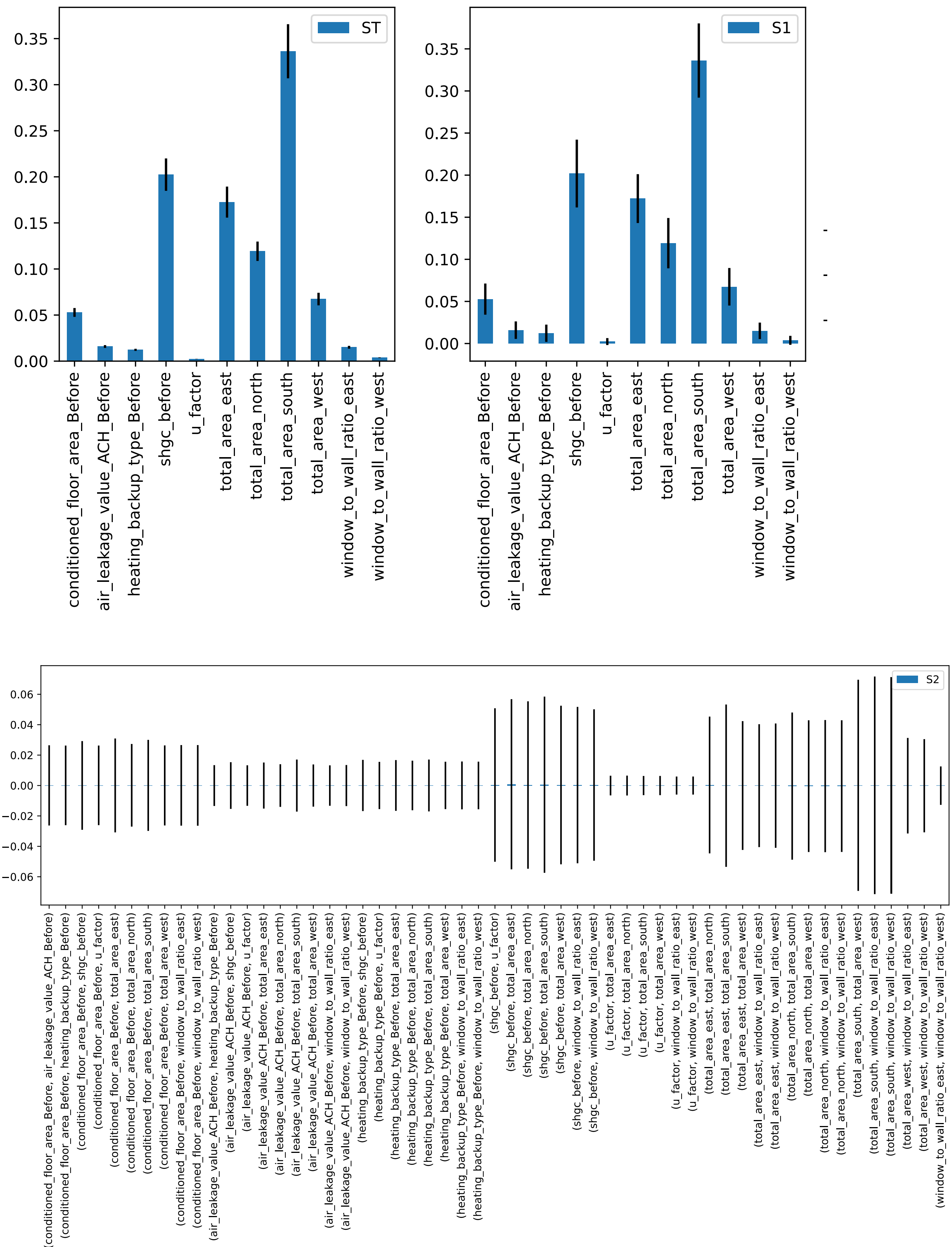

**Fig. 8.** Sobol sensitivity analysis plot of Study 1B: increasing south window SHGC to 0.7 for heat-pump-upgraded Chicago residential building stock.

Most influential variables for emission savings are the total window area on the south façade, SHGC before the retrofit, and the total window area on the east façade.

**Study 2A** — decreasing all window U-value to 2.56 $W/m^2{\cdot}K$ and increasing all window SHGC to 0.7 for the 2018 residential building stock of

Chicago:

Initial OLS linear regression model results (includes every relevant variable):

R-squared is 0.370.

| Variable | Coef | Std Err | t | P>\|t\| | [0.025 | 0.975] |
|---|---|---|---|---|---|---|
| const | −1.4E + 10 | 8.37E + 10 | −0.169 | 0.865 | −1.8E + 11 | 1.5E + 11 |
| in.tenure_Metadata | −0.0369 | 0.015 | −2.435 | 0.015 | −0.067 | −0.007 |
| in.usage_level_Metadata | −0.0048 | 0.017 | −0.274 | 0.784 | −0.039 | 0.029 |
| conditioned_floor_area_Before | −0.3689 | 0.064 | −5.8 | 0 | −0.494 | −0.244 |
| air_leakage_value_ACH_Before | 0.7524 | 0.066 | 11.376 | 0 | 0.623 | 0.882 |
| heating_backup_efficiency_value_Before | −0.9253 | 0.218 | −4.24 | 0 | −1.353 | −0.497 |
| cooling_primary_efficiency_value | −0.0907 | 0.068 | −1.339 | 0.181 | −0.223 | 0.042 |
| year_normalized | −0.0235 | 0.023 | −1.029 | 0.303 | −0.068 | 0.021 |
| income_normalized | −0.042 | 0.021 | −1.983 | 0.047 | −0.084 | 0 |
| roof_assembly_effective_rvalue | −0.0517 | 0.023 | −2.228 | 0.026 | −0.097 | −0.006 |
| average_floor_rvalue | 0.0433 | 0.04 | 1.082 | 0.279 | −0.035 | 0.122 |
| shgc_before | −0.6585 | 0.083 | −7.899 | 0 | −0.822 | −0.495 |
| u_factor | 1.2199 | 0.058 | 20.892 | 0 | 1.105 | 1.334 |
| projection_factor_east | −0.1383 | 0.032 | −4.341 | 0 | −0.201 | −0.076 |
| projection_factor_north | −0.1235 | 0.03 | −4.121 | 0 | −0.182 | −0.065 |
| projection_factor_south | −0.1421 | 0.031 | −4.599 | 0 | −0.203 | −0.082 |
| projection_factor_west | −0.1401 | 0.031 | −4.534 | 0 | −0.201 | −0.08 |
| total_area_east | −0.1762 | 0.189 | −0.934 | 0.35 | −0.546 | 0.193 |
| total_area_north | −0.1271 | 0.195 | −0.652 | 0.514 | −0.509 | 0.255 |
| total_area_south | −0.075 | 0.23 | −0.327 | 0.744 | −0.525 | 0.375 |
| total_area_west | 0.0764 | 0.186 | 0.411 | 0.681 | −0.288 | 0.441 |
| window_to_wall_ratio_east | 0.2284 | 0.052 | 4.412 | 0 | 0.127 | 0.33 |
| window_to_wall_ratio_north | 0.2874 | 0.052 | 5.519 | 0 | 0.185 | 0.389 |
| window_to_wall_ratio_south | 0.2192 | 0.052 | 4.243 | 0 | 0.118 | 0.321 |
| window_to_wall_ratio_west | 0.1509 | 0.051 | 2.987 | 0.003 | 0.052 | 0.25 |
| residential_facility_type_Before_apartment unit | 1.07E + 10 | 6.33E + 10 | 0.169 | 0.865 | −1.1E + 11 | 1.35E + 11 |
| residential_facility_type_Before_single-family attached | 1.07E + 10 | 6.33E + 10 | 0.169 | 0.865 | −1.1E + 11 | 1.35E + 11 |
| residential_facility_type_Before_single-family detached | 1.07E + 10 | 6.33E + 10 | 0.169 | 0.865 | −1.1E + 11 | 1.35E + 11 |
| heating_backup_fuel_Before_electricity | 2.24E + 10 | 1.32E + 11 | 0.169 | 0.865 | −2.4E + 11 | 2.82E + 11 |
| heating_backup_fuel_Before_fuel oil | 2.24E + 10 | 1.32E + 11 | 0.169 | 0.865 | −2.4E + 11 | 2.82E + 11 |
| heating_backup_fuel_Before_natural gas | 2.24E + 10 | 1.32E + 11 | 0.169 | 0.865 | −2.4E + 11 | 2.82E + 11 |
| heating_backup_fuel_Before_propane | 2.24E + 10 | 1.32E + 11 | 0.169 | 0.865 | −2.4E + 11 | 2.82E + 11 |
| heating_primary_type_Boiler | −1E + 10 | 6.15E + 10 | −0.169 | 0.865 | −1.3E + 11 | 1.1E + 11 |
| heating_primary_type_ElectricResistance | −1E + 10 | 6.15E + 10 | −0.169 | 0.865 | −1.3E + 11 | 1.1E + 11 |
| heating_primary_type_Furnace | −1E + 10 | 6.15E + 10 | −0.169 | 0.865 | −1.3E + 11 | 1.1E + 11 |
| heating_primary_type_HeatPump | −9.5E + 09 | 5.6E + 10 | −0.169 | 0.865 | −1.2E + 11 | 1E + 11 |
| heating_primary_type_WallFurnace | −1E + 10 | 6.15E + 10 | −0.169 | 0.865 | −1.3E + 11 | 1.1E + 11 |
| cooling_primary_type_0 | −8.5E + 09 | 5.04E + 10 | −0.169 | 0.865 | −1.1E + 11 | 9.02E + 10 |
| cooling_primary_type_HeatPump | −9.5E + 09 | 5.59E + 10 | −0.169 | 0.865 | −1.2E + 11 | 1E + 11 |
| cooling_primary_type_central air conditioner | −8.5E + 09 | 5.04E + 10 | −0.169 | 0.865 | −1.1E + 11 | 9.02E + 10 |
| cooling_primary_type_mini-split | −8.5E + 09 | 5.04E + 10 | −0.169 | 0.865 | −1.1E + 11 | 9.02E + 10 |
| cooling_primary_type_room air conditioner | −8.5E + 09 | 5.04E + 10 | −0.169 | 0.865 | −1.1E + 11 | 9.02E + 10 |

Refitted OLS model (removing P>|t| less than 0.05):

R-squared is 0.353.

| Variable | Coef | Std Err | t | P>\|t\| | [0.025 | 0.975] |
|---|---|---|---|---|---|---|
| const | 0.2911 | 0.038 | 7.628 | 0 | 0.216 | 0.366 |
| in.tenure_Metadata | −0.0312 | 0.014 | −2.295 | 0.022 | −0.058 | −0.005 |
| conditioned_floor_area_Before | −0.3809 | 0.06 | −6.373 | 0 | −0.498 | −0.264 |
| air_leakage_value_ACH_Before | 0.7973 | 0.065 | 12.258 | 0 | 0.67 | 0.925 |
| heating_backup_efficiency_value_Before | −0.4265 | 0.046 | −9.215 | 0 | −0.517 | −0.336 |
| income_normalized | −0.0514 | 0.021 | −2.464 | 0.014 | −0.092 | −0.01 |
| roof_assembly_effective_rvalue | −0.0534 | 0.023 | −2.331 | 0.02 | −0.098 | −0.008 |
| shgc_before | −0.6728 | 0.084 | −8.005 | 0 | −0.838 | −0.508 |
| u_factor | 1.2318 | 0.059 | 20.946 | 0 | 1.117 | 1.347 |
| projection_factor_east | −0.1197 | 0.031 | −3.924 | 0 | −0.179 | −0.06 |
| projection_factor_north | −0.1045 | 0.029 | −3.659 | 0 | −0.161 | −0.049 |
| projection_factor_south | −0.1206 | 0.029 | −4.119 | 0 | −0.178 | −0.063 |
| projection_factor_west | −0.1259 | 0.029 | −4.284 | 0 | −0.183 | −0.068 |
| window_to_wall_ratio_east | 0.1832 | 0.028 | 6.441 | 0 | 0.127 | 0.239 |
| window_to_wall_ratio_north | 0.2573 | 0.028 | 9.071 | 0 | 0.202 | 0.313 |
| window_to_wall_ratio_south | 0.1951 | 0.028 | 7.066 | 0 | 0.141 | 0.249 |
| window_to_wall_ratio_west | 0.1644 | 0.027 | 6.001 | 0 | 0.111 | 0.218 |

Top 3 coefficients from refitted OLS model (largest absolute value):

u_factor, air_leakage_value_ACH_Before, shgc_before.

Result of Sobol sensitivity analysis:

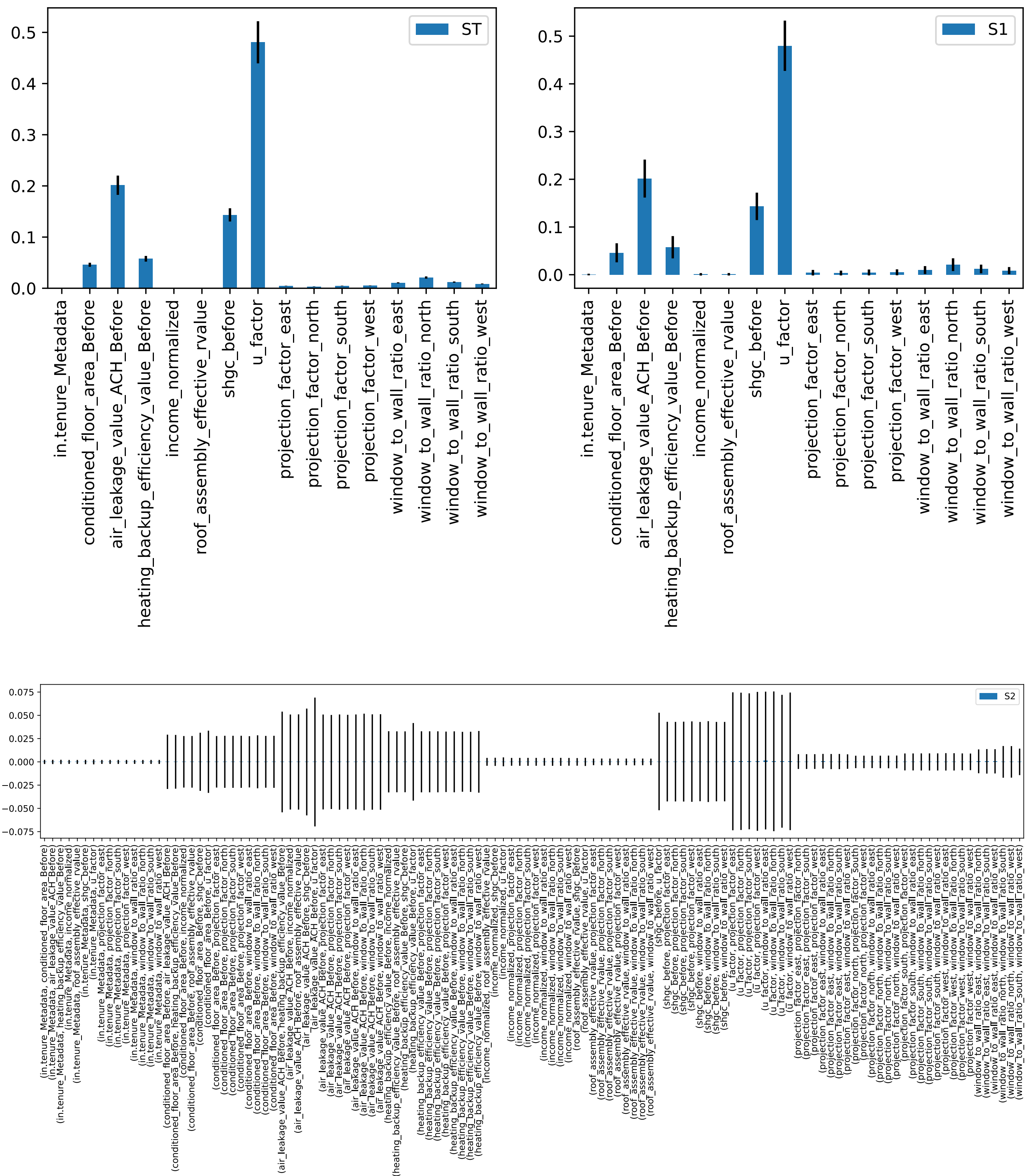

**Fig. 9.** Sobol sensitivity analysis plot of Study 2A: decreasing all window U-value to 2.56 W/m$^2$·K and increasing all window SHGC to 0.7 for the 2018 residential building stock of Chicago.

Most influential variables for emission savings are the initial U-value, air leakage value, and SHGC before the retrofit.

**Study 2B** — decreasing all window U-value to 2.56 W/m$^2$·K and increasing all window SHGC to 0.7 for heat-pump-upgraded Chicago residential building stock**:**

Initial OLS linear regression model results (includes every relevant variable):

R-squared is 0.522.

| Variable | Coef | Std Err | t | P>\|t\| | [0.025 | 0.975] |
|---|---|---|---|---|---|---|
| const | −2E + 10 | 2.56E + 10 | −0.782 | 0.434 | −7E + 10 | 3.02E + 10 |
| in.tenure_Metadata | −0.0158 | 0.009 | −1.801 | 0.072 | −0.033 | 0.001 |
| in.usage_level_Metadata | −0.0039 | 0.01 | −0.4 | 0.689 | −0.023 | 0.015 |
| conditioned_floor_area_Before | −0.1889 | 0.039 | −4.79 | 0 | −0.266 | −0.112 |
| air_leakage_value_ACH_Before | 0.2887 | 0.038 | 7.523 | 0 | 0.213 | 0.364 |
| heating_backup_type_Before | 0.227 | 0.009 | 26.65 | 0 | 0.21 | 0.244 |
| heating_backup_efficiency_value_Before | −0.0341 | 0.023 | −1.456 | 0.145 | −0.08 | 0.012 |
| year_normalized | −0.0301 | 0.013 | −2.276 | 0.023 | −0.056 | −0.004 |
| income_normalized | −0.0192 | 0.012 | −1.548 | 0.122 | −0.044 | 0.005 |
| roof_assembly_effective_rvalue | −0.0262 | 0.013 | −1.984 | 0.047 | −0.052 | 0 |
| average_floor_rvalue | −0.0312 | 0.028 | −1.109 | 0.267 | −0.086 | 0.024 |
| shgc_before | −0.3039 | 0.047 | −6.466 | 0 | −0.396 | −0.212 |
| u_factor | 0.8658 | 0.033 | 26.283 | 0 | 0.801 | 0.93 |
| projection_factor_east | −0.0478 | 0.018 | −2.655 | 0.008 | −0.083 | −0.013 |
| projection_factor_north | −0.0731 | 0.017 | −4.313 | 0 | −0.106 | −0.04 |
| projection_factor_south | −0.0576 | 0.017 | −3.301 | 0.001 | −0.092 | −0.023 |
| projection_factor_west | −0.0856 | 0.017 | −4.9 | 0 | −0.12 | −0.051 |
| total_area_east | −0.0492 | 0.106 | −0.462 | 0.644 | −0.258 | 0.16 |
| total_area_north | 0.0407 | 0.112 | 0.365 | 0.715 | −0.178 | 0.259 |
| total_area_south | 0.1021 | 0.13 | 0.787 | 0.431 | −0.152 | 0.356 |
| total_area_west | −0.2306 | 0.106 | −2.175 | 0.03 | −0.439 | −0.023 |
| window_to_wall_ratio_east | 0.1133 | 0.029 | 3.868 | 0 | 0.056 | 0.171 |
| window_to_wall_ratio_north | 0.1374 | 0.03 | 4.624 | 0 | 0.079 | 0.196 |
| window_to_wall_ratio_south | 0.1406 | 0.029 | 4.82 | 0 | 0.083 | 0.198 |
| window_to_wall_ratio_west | 0.1384 | 0.029 | 4.811 | 0 | 0.082 | 0.195 |
| residential_facility_type_Before_apartment unit | 1.08E + 11 | 1.38E + 11 | 0.782 | 0.434 | −1.6E + 11 | 3.77E + 11 |
| residential_facility_type_Before_single-family attached | 1.08E + 11 | 1.38E + 11 | 0.782 | 0.434 | −1.6E + 11 | 3.77E + 11 |
| residential_facility_type_Before_single-family detached | 1.08E + 11 | 1.38E + 11 | 0.782 | 0.434 | −1.6E + 11 | 3.77E + 11 |
| heating_backup_fuel_Before_electricity | −8.8E + 10 | 1.12E + 11 | −0.782 | 0.434 | −3.1E + 11 | 1.32E + 11 |
| heating_backup_fuel_Before_fuel oil | −8.8E + 10 | 1.12E + 11 | −0.782 | 0.434 | −3.1E + 11 | 1.32E + 11 |
| heating_backup_fuel_Before_natural gas | −8.8E + 10 | 1.12E + 11 | −0.782 | 0.434 | −3.1E + 11 | 1.32E + 11 |
| heating_backup_fuel_Before_propane | −8.8E + 10 | 1.12E + 11 | −0.782 | 0.434 | −3.1E + 11 | 1.32E + 11 |

Refitted OLS model (removing P>|t| less than 0.05):

R-squared is 0.480.

| Variable | Coef | Std Err | t | P>\|t\| | [0.025 | 0.975] |
|---|---|---|---|---|---|---|
| const | −0.1411 | 0.022 | −6.495 | 0 | −0.184 | −0.099 |
| conditioned_floor_area_Before | −0.2123 | 0.033 | −6.462 | 0 | −0.277 | −0.148 |
| air_leakage_value_ACH_Before | 0.358 | 0.038 | 9.459 | 0 | 0.284 | 0.432 |
| heating_backup_type_Before | 0.2013 | 0.007 | 27.516 | 0 | 0.187 | 0.216 |
| roof_assembly_effective_rvalue | −0.0341 | 0.013 | −2.586 | 0.01 | −0.06 | −0.008 |
| shgc_before | −0.3358 | 0.049 | −6.88 | 0 | −0.431 | −0.24 |
| u_factor | 0.8784 | 0.034 | 25.706 | 0 | 0.811 | 0.945 |
| projection_factor_north | −0.058 | 0.017 | −3.503 | 0 | −0.091 | −0.026 |
| projection_factor_south | −0.0428 | 0.017 | −2.523 | 0.012 | −0.076 | −0.01 |
| projection_factor_west | −0.0443 | 0.017 | −2.598 | 0.009 | −0.078 | −0.011 |
| window_to_wall_ratio_east | 0.0952 | 0.015 | 6.292 | 0 | 0.066 | 0.125 |
| window_to_wall_ratio_north | 0.1393 | 0.016 | 8.452 | 0 | 0.107 | 0.172 |
| window_to_wall_ratio_south | 0.1548 | 0.016 | 9.672 | 0 | 0.123 | 0.186 |
| window_to_wall_ratio_west | 0.0769 | 0.016 | 4.899 | 0 | 0.046 | 0.108 |

Top 3 coefficients from refitted OLS model (largest absolute value):

u_factor, air_leakage_value_ACH_Before, shgc_before.

Result of Sobol sensitivity analysis:

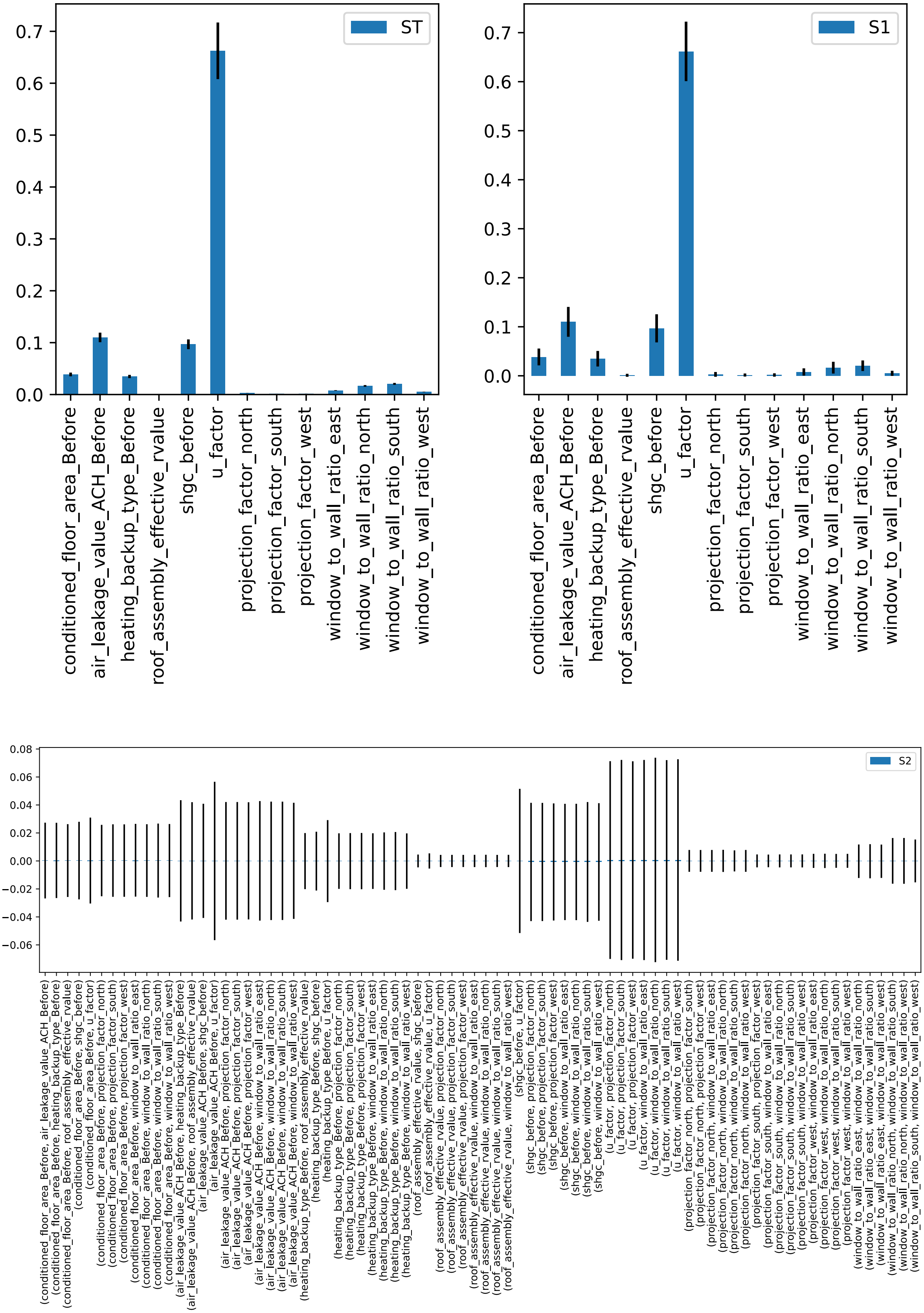

**Fig. 10.** Sobol sensitivity analysis plot of Study 2B: decreasing all window U-value to 2.56 W/m$^2$·K and increasing all window SHGC to 0.7 for heat-pump-upgraded Chicago residential building stock.

Most influential variables for emission savings are the initial U-value, air leakage value, and SHGC before the retrofit.

## Data availability

Data will be made available on request.